\documentclass[useAMS, usenatbib]{mn2e}
\usepackage{graphicx, rotate}
\usepackage{natbib, lscape}
\usepackage{latexsym, amssymb, verbatim}
\usepackage{eufrak}
\usepackage{amsmath}
\usepackage{epsfig}
\usepackage{times}
\usepackage{rotating}
\usepackage{graphicx}
\usepackage{color}
\usepackage{psfrag}

\title[\sc torus-3dpdr]
{{\sc torus-3dpdr}: A self-consistent code treating three-dimensional photoionization and photodissociation regions}
\author[T. G. Bisbas et al.]{T. G. Bisbas$^{1,2,3,7}$\thanks{E-mail: tb@mpe.mpg.de}, T. J. Haworth$^{4,7}$, M. J. Barlow$^{1}$, S. Viti$^{1}$, T. J. Harries$^{5}$, T. Bell$^{6}$, \newauthor and J. A. Yates$^{1}$\\
$^{1}$ Department of Physics and Astronomy, Kathleen Lonsdale Building, University College London, Gower Street, London WC1E 6BT, U.K.\\
$^{2}$ Max-Planck-Institut f\"ur Extraterrestrische Physik, Giessenbachstrasse 1, D-85748 Garching, Germany\\
$^{3}$ Department of Astronomy and Physics, University of Florida, Gainesville, FL 32611, USA\\
$^{4}$ Institute of Astronomy, University of Cambridge, Madingley Road, Cambridge CB3 0HA, UK\\
$^{5}$ School of Physics, University of Exeter, Stocker Road, Exeter EX4 4QL\\
$^{6}$ Centro de Astrobiolog\'ia (CSIC-INTA), Torrej\'on de Ardoz, 28850 Madrid, Spain\\
$^{7}$ NORDITA, KTH Royal Institute of Technology and Stockholm University, Roslagstullsbacken 23, 10691 Stockholm, Sweden}
\date{Accepted , Received ; in original form \today}

\begin{document}

\maketitle

\begin{abstract}
The interaction of ionizing and far-ultraviolet radiation with the interstellar medium is of great importance. It results in the formation of regions in which the gas is ionized, beyond which are photodissociation regions (PDRs) in which the gas transitions to its atomic and molecular form. Several numerical codes have been implemented to study these two main phases of the interstellar medium either dynamically or chemically. In this paper we present {\sc torus-3dpdr}, a new self-consistent code for treating the chemistry of three-dimensional photoionization and photodissociation regions. It is an integrated code coupling the two codes {\sc torus}, a hydrodynamics and Monte Carlo radiation transport code, and {\sc 3d-pdr}, a photodissociation regions code. The new code uses a Monte Carlo radiative transfer scheme to account for the propagation of the ionizing radiation including the diffusive component as well as a ray-tracing scheme based on the {\sc healpix} package in order to account for the escape probability and column density calculations. Here, we present the numerical techniques we followed and we show the capabilities of the new code in modelling three-dimensional objects including single or multiple sources. We discuss the effects introduced by the diffusive component of the UV field in determining the thermal balance of PDRs as well as the effects introduced by a multiple sources treatment of the radiation field. We find that diffuse radiation can positively contribute to the formation of CO. With this new code, three-dimensional synthetic observations for the major cooling lines are possible, for making feasible a detailed comparison between hydrodynamical simulations and observations.
\end{abstract}

\begin{keywords}
ISM: abundances -- astrochemistry -- radiative transfer -- methods: numerical
\end{keywords}

%%%%%%%%%%%%%%%%%%%%%%%
\section{Introduction}%
%%%%%%%%%%%%%%%%%%%%%%%

The interstellar medium (ISM) consists predominantly of hydrogen ($70\%$), followed by helium ($28\%$) and heavier elements ($2\%$). Its interaction with the ultraviolet (UV) radiation emitted by massive stars is of great interest. Extreme UV photons carrying more energy than the ionization potential of hydrogen ($13.6\,{\rm eV}$) ionize the gas. Any excess over the ionization potential is then transformed into kinetic energy of the liberated electrons (photoelectrons). Regions containing a significantly high fraction of photoelectrons are known as `H~{\sc ii} regions' \citep{Stro39} and have temperatures of the order of $5,000-15,000\,{\rm K}$ which is higher than that of the surrounding gas ($\le100\,{\rm K}$) resulting in their expansion \citep{Kahn54,Spit78}. The edge of the ionized medium is known as an `ionization front' and separates the H~{\sc ii} region from the rest of the ISM. The ionization front can be considered as a sharp discontinuity in which the degree of ionization and the gas temperature drops abruptly, over a few UV photon mean free paths. The material ahead of this front is dominated by far-UV radiation (FUV) carrying energy $6<h\nu<13.6\,{\rm eV}$. Photons in this energy regime can photodissociate CO and H$_2$ molecules creating so called `photodissociation regions' \citep[PDRs; known also as `photon-dominated regions', see][for a review]{Holl99}. Further ahead the radiation field is attenuated to the extent that it does not affect quiescent molecular zones. 

Studies of the different phases of the ISM are of great interest as they can lead to an in-depth understanding of the mechanisms responsible for its structure and dynamical evolution. This in turn may answer important questions concerning the contribution of stellar feedback in driving turbulence at small (subpc) scales \citep[see e.g.][for a more general discussion on ISM turbulence]{Kles14} well as the modes that trigger the formation of new stars beyond the expanding ionized regions. Studying PDRs is key to understanding the role played by the FUV photons in governing the physical and chemical structure and determining the thermal balance of the neutral ISM in galaxies.

Several groups worldwide have made significant efforts in simulating the (hydro-)dynamical evolution of the ISM using different computational techniques such as Smoothed Particle Hydrodynamics (SPH) and Adaptive Mesh Refinement (AMR). These have led to the modelling of star formation processes, as well as feedback from massive stars and magnetic fields \citep{Hubb11, Hubb13, Bisb09, Bisb11, Walc12, Walc13, Dale05, Dale07, Dale13, Grit09a, Grit09b, Arth11, Mell06}. On the other hand, observations have produced a wealth of data from the atomic and molecular lines emitted at different wavelengths by the different phases of the ISM. However, this does not directly reveal the exact three-dimensional density and velocity structure of the observed objects. There is thus the need for a tool which acts as a ``common interface'' between the simulated hydrodynamical ISM structure and the observed line emission. Towards this goal, efforts have been made to model the complex chemical processes occurring in the different phases of the ISM of particular relevance to PDRs. Coupling radiation transport techniques with hydrodynamics and the gas chemistry, however, is complicated and poorly developed as yet, although significant efforts have been made in this direction by \citet{Glov07, Glov07b, Dobb08, Glov10} and \citet{Levr12}.

Numerical codes dedicated to modelling the chemistry in photoionized regions and PDRs have been presented in the past. Codes for photoionized regions include {\sc cloudy} \citep{Ferl98} while a three-dimensional treatment has been achieved in the {\sc mocassin} code \citep{Erco03, Wood04, Erco05} and in the {\sc torus} code \citep{Harr00,Hawo12} which are based on Monte Carlo (MC) methods first described by \citet{Lucy99}. On the other hand, several codes treating PDRs have been presented in the past decade or so. These include {\sc ucl\_pdr} \citep{Papa02, Bell05, Bell06}, {\sc cloudy} \citep{Ferl98,Abel05,Shaw05}, {\sc costar} \citep{Kamp00,Kamp01}, {\sc htbkw} \citep{Tiel85,Kauf99,Wolf03}, {\sc kosma}-$\tau$ \citep{Stoe96,Bens03,Roll06}, {\sc leiden} \citep{Blac87b,vanD88,Jans95}, {\sc meijerink} \citep{Meij05}, {\sc meudon} \citep{LePe04, LePe02, LeBo93} and {\sc sternberg} \citep{Ster89, Ster95, Boge06}. 

\citet{Roll07} provide an interesting review of many of those PDR codes as well as benchmarking tests pinpointing the differences. The codes are able to treat complicated chemical networks and thus obtain the abundances of many chemical species, the line emissivities of different coolants, as well as temperature profiles within a given PDR. While most attention has been paid to an accurate chemical PDR modelling, the majority of PDR codes treat one-dimensional structures. Recently \citet{Bisb12} have implemented the three-dimensional code {\sc 3d-pdr} which is able to treat PDRs of any given arbitrary density distribution. In addition \citet{Andr14} also improved the {\sc kosma}-$\tau$ PDR code to handle 3D pixels (voxels) which mimic the fractal structure of the ISM. These new implementations meet the need for PDR codes to treat complicated geometrical structures, since the chemistry can depend on it.

In this paper, we present the first self-consistent fully three-dimensional unified code for treating photoionized and photodissociated regions simultaneously with arbitrary geometrical and density distributions. This new code is able to calculate i) the structure of circumstellar and interstellar radiation fields using an MC method that treats the diffuse component and ii) a detailed PDR chemistry using a {\sc healpix} ray-tracing scheme to estimate the cooling and heating rates as well as line emissivities, the abundances of chemical species and temperature profiles, following the methods of the {\sc 3d-pdr} code. It therefore models the wide range of observable chemical processes which take place in the transition zones from ionized to molecular regions. 

Our paper is organized as follows. In Sections \ref{sec:3dpdr} and \ref{sec:torus} we give an overview of the {\sc 3d-pdr} and {\sc torus} codes respectively. In Section \ref{sec:coupling} we show the method we followed for coupling the two codes. In Section \ref{sec:benchmark} we benchmark our new code while in Section \ref{sec:applications} we show the capabilities of the code in simulating three-dimensional structures. We conclude in Section \ref{sec:conclusions} with the description of the effects introduced in simulating photoionized and photodissociation regions in full three dimensions.

%%%%%%%%%%%%%%%%%%%%%%%%%%%%%%%%
\section{The {\sc 3d-pdr} code}%
%%%%%%%%%%%%%%%%%%%%%%%%%%%%%%%%
\label{sec:3dpdr}

The {\sc 3d-pdr} code \citep[][ hereafter `B12']{Bisb12} is a three-dimensional time-dependent astrochemistry code which has been designed to treat PDRs of arbitrary density distribution. It is a further development of the one-dimensional {\sc ucl\_pdr} code and it uses the chemical model features of \citet{Bell06}. It solves the chemistry and the thermal balance self-consistently in each computational element of a given cloud. {\sc 3d-pdr} has been fully benchmarked against other PDR codes according to the tests of \citet{Roll07} and has been used in various applications e.g. \citet{Offn13, Offn14, Bisb14, Bisb15, Gach15}. 

We list below a brief overview emphasizing on the ray-tracing and UV treatment as well as the model chemistry used in {\sc 3d-pdr}. For full details about the code see B12.

\subsection{Ray-tracing and UV treatment}
{\sc 3d-pdr} uses a ray-tracing scheme based on the {\sc healpix} \citep{Gors05} package. Along each ray, the elements closest to the line-of-sight are projected creating a set of points called `evaluation points'. The properties of the evaluation points are identical to those of the projected elements. With this technique we are able to calculate: i) the column densities of species for a random element along a particular direction, ii) the attenuation of the far-ultraviolet radiation in the PDR, and iii) the propagation of the far-infrared/submm line emission out of the PDR.

The treatment of the UV radiation in the {\sc 3d-pdr} code is initially estimated by invoking the {\it on-the-spot} approximation i.e. by neglecting the diffuse component of the radiation field. The attenuation of the UV field, $\chi$, at a randomly given point, $p$, is evaluated using the equation
\begin{eqnarray}
\label{eqn:3dpdrUV}
\chi(p)=\frac{1}{{\cal N}_{\ell}}\sum_{{\bf q}=1}^{{\cal N}_{\ell}}\chi_o({\bf q})e^{-\tau_{\rm UV}A_{\rm V}({\bf q})}\,,
\end{eqnarray}
where $\chi_o({\bf q})$ is the magnitude of the unattenuated field strength along the {\sc healpix} direction ${\bf q}$, ${\cal N}_{\ell}=12\times4^{\ell}$ is the number of {\sc healpix} rays at level of refinement $\ell$, $\tau_{\rm UV}=3.02$ is a dimensionless factor converting the visual extinction to UV attenuation \citep{Bell06}, and $A_{\rm V}$ is the visual extinction along ${\bf q}$ defined as
\begin{eqnarray}
\label{eqn:AV}
A_{\rm V}({\bf q})=A_{\rm V,o}\int_0^Ln_{\rm H}dr\,.
\end{eqnarray}
In the above equation $A_{\rm V,o}=5.3\times10^{-22}\,{\rm mag}\,{\rm cm}^{2}$ and the integration corresponds to the column of the H-nucleus number density $n_{\rm H}$ along the ray of length $L$. We note that {\sc torus-3dpdr} performs radiative transfer with frequency dependent opacity, however for the purposes of this initial comparison in the present work and consistency with the majority of the PDR codes we have restricted $\tau_{\rm UV}$ to the above value \citep[c.f.][]{Pint09}.

Equation \ref{eqn:3dpdrUV} has been extensively used in several codes treating PDRs. As described in \citet{Roll07}, most of the models adopt a plane-parallel geometry illuminated from one or from both sides. However, the majority of them do not account for the geometrical dilution i.e. that a spherically emitted radiation field decreases in intensity with the square of the distance. Accounting for the latter, the attenuation along a {\sc healpix} direction $\bf q$ is expressed as follows:
\begin{eqnarray}
\label{eqn:geomdil}
\chi(r,{\bf q})=\chi_{\rm o}({\bf q})e^{-\tau_{\rm UV}A_{\rm V}({\bf q})}\frac{R_{_{\rm IF}}^2}{(R_{_{\rm IF}}+r)^2},
\end{eqnarray}
where $R_{_{\rm IF}}$ is the position of the ionization front from the exciting source, and $r$ is the distance of a particular point $p$ from the ionization front and inside the PDR (see Appendix \ref{app:dilution} for the corresponding derivation). As we explain below, this is an important factor especially when a PDR model is being used to reproduce the observed data of an object. In {\sc torus-3dpdr} this attenuation naturally comes from the Monte Carlo (MC) radiative transfer treatment and this is what is being used throughout this paper unless otherwise stated.

\subsection{Model chemistry}

The code uses the most recent UMIST 2012 chemical network database \citep{McEl13}. This network consists of 215 species and more than 3000 reactions. However, in this paper we will consider only a subset of this network consisting of 33 species (including ${\rm e}^{-}$) and 330 reactions. We make use of the {\sc sundials} package\footnote{http://computation.llnl.gov/casc/sundials/main.html} in order to construct the appropriate set of ordinary differential equations (ODEs) for describing the formation and destruction of each species as well as the associated Jacobian matrices to compute the abundance of each chemical species. 

For the particular cases of H$_2$ and CO photodissociation rates, we adopt the treatments of \citet{Lee96} and \citet{vanD88} and we use the additional tabulated shielding functions provided in those papers. We also account for the shielding of C~{\sc i} using the treatment of \citet{Kamp00} in order to estimate the photoionization rate of carbon. The rate of molecular hydrogen formation on dust grains is calculated using the treatment of \citet{Caza02,Caza04} while the thermally averaged sticking coefficient of hydrogen atoms on dust grains is taken from \citet{Holl79}. The dust temperature at each point in the density distribution is calculated using the treatment of \citet{Holl91} to account for the grain heating due to the incident FUV photons. Future updates will include more detailed heating functions such as \citet{Wein01} and \citet{Comp11} for the grain charge photo-electric heating and grain temperature respectively.

%%%%%%%%%%%%%%%%%%%%%%%%%%%%%%%
\section{The {\sc torus} code}%
%%%%%%%%%%%%%%%%%%%%%%%%%%%%%%%
\label{sec:torus}

\textsc{torus} is a grid-based three-dimensional radiation transport and  Eulerian hydrodynamics code that uses an octree adaptive (non-uniform) grid. 

It was initially designed to perform three-dimensional calculations, including the treatment of polarization and Mie or Rayleigh scattering. In this form it was used to analyze synthetic spectral line observations of stellar winds that are rotationally distorted by rapid stellar rotation or which contain clumps \citep{Harr00}. With the addition of a dust treatment it was also used to model observations of the Wolf-Rayet (WR) star binary WR137 in an effort to provide an explanation for observed polarization variability \citep{Harr00b}. It has since continued developing and been applied to models of accretion on to T Tauri stars \citep{Vink05,Symi05}, discs around Herbig AeBe stars \citep{Tann08}, Raman-scattered line formation in symbiotic binaries \citep{Harr97}, dust emission and molecular line formation in star forming regions \citep{Kuro04,Rund10} and synthetic galactic HI observations \citep{Acre10b}. Most recently it has been used in radiation hydrodynamic applications, both by coupling it to an SPH code \citep{Acre10} and in a self-contained manner by performing the hydrodynamics calculation on the \textsc{torus} grid \citep{Hawo12}.

We give below a brief overview emphasizing on the key features of the MC photoionization routines.

\subsection{Monte Carlo Photoionization}
\label{ssec:mcphot}

\textsc{torus} performs photoionization calculations using an iterative MC photon energy packet propagating routine, similar to that of \cite{Erco03} and \cite{Wood04} which in turn are based on the methods presented by \cite{Lucy99}. These photon energy packets are collections of photons for which the total energy $\epsilon$ remains constant, but the number of photons contained varies for different frequencies $\nu$. In the model, they are initiated at stars with frequencies selected randomly based on the emission spectrum of the star. The constant energy value $\epsilon$ for each photon packet is simply the total energy emitted by star's (luminosity $L$) during the duration $\Delta t$ of the iteration divided by the total number of photon packets $N$:
\begin{equation}
\epsilon = \frac{L \Delta t}{N}.
\end{equation}
Photons that are emitted from a given ionizing source are propagated in random but isotropic directions. As soon as a photon packet is emitted, it will propagate for a path length $l$ determined by a randomly selected optical depth. The length $l$ is determined by the position in which the next event will occur; either involving an interaction with the material after traversing a random optical depth given by
\begin{equation}
	\tau = - \ln(1-r)\,;\,\,r\in[0,1),
\end{equation}
\citep[as detailed in][]{Harr97}, or involving the crossing of a cell boundary.

If the photon packet fails to escape a cell after traveling an optical depth $\tau$ then its propagation ceases and an absorption event occurs: this is the situation when $\tau>\tau_{\rm CELL}$, $\tau_{\rm CELL}=\kappa\cdot dr$ where $\kappa$ is the opacity term and $dr$ the size of the cell \citep[see also][]{Erco03}. At this point, there are two possibilities based on the assumed effect of the diffuse field radiation:
\begin{enumerate}
\item{Accounting for the diffuse field. The diffuse field refers to photons emitted by the recombination of electrons with ions. The effect of this radiation field is both to further ionize the gas (should the recombination photon be energetic enough) and to cool the gas (through extraction of energy by optically thin photons). The diffuse radiation field can be important for irradiating regions shadowed by denser clumps along their line-of-sight with the ionizing source. Here, the total recombination coefficient (so-called `case-A') is given by
\begin{eqnarray}
\alpha_{\rm A}=\sum_{n=1}^{\infty}\alpha_{\rm n}(H^{\circ},T)
\end{eqnarray}
where $\alpha_{\rm n}(H^{\circ},T)$ is the recombination coefficient of hydrogen at temperature $T$ in the quantum level $n$. \citet{Vern96a} studied how $\alpha_{\rm A}$ changes by varying the temperature for the basic elements (H, He, Li, and Na). In {\sc torus-3dpdr}, once an absorption event occurs it it is followed by the immediate emission of a new photon packet from the same location and with the same energy, but containing a different number of photons to account for different photon frequency. We also note that the diffuse component of radiation field can be created by the scattering of ionizing photons onto dust grains. Such treatment has been performed e.g. \citet{Erco05} and we aim to include the dust contribution in a subsequent paper.} 
\item Using the {\it on-the-spot} \citep[OTS; ][]{Oste89} approximation (case-B) in which the diffuse field photons are assumed to contribute negligibly to the global ionization structure following absorption. The recombination coefficient is then given by equation
\begin{eqnarray}
\alpha_{\rm B}=\alpha_{\rm A} - \alpha_{1}(H^{\circ},T)=\sum_{n=2}^{\infty}\alpha_{\rm n}(H^{\circ},T)
\end{eqnarray}
This is justified in regions of simple geometry, for example where density gradients are small. In MC photoionization with the OTS approximation, once a photon packet is absorbed it is ignored and is assumed to either have been re-emitted with a frequency lower than that required for photoionization, or to provide a negligible further contribution to the ionization structure by causing further photoionization on only small scales. 
\end{enumerate}
 
The energy density $dU$ of a radiation field is given by
\begin{equation}
	dU = \frac{4\pi J_{\nu}}{c} d \nu,
\label{energydensity}
\end{equation}
where $c$ is the speed of light and $J_{\nu}$ is the specific intensity and frequency $\nu$. A photon energy packet traversing a path $l$ in a particular cell contributes an energy $\epsilon (l/c) / \Delta t$ to the time-averaged energy density of that cell. Thus by summing over all paths $l$ the energy density of a given cell (volume $V$) can be determined. Thus Eqn.~\ref{energydensity} can be evaluated using the expression:
\begin{equation}
	\frac{4\pi J_{\nu}}{c} d \nu = \frac{\epsilon}{c\Delta t}
        \frac{1}{V} \sum_{d\nu} l.
\label{energydensityMC}
\end{equation}
This is then used to obtain ionization fractions by solving the ionization balance equation \citep{Oste89}
\begin{equation}
	\frac{n(X^{i+1})}{n(X^i)} = \frac{1}{\alpha(X^i) n_e} \int_{\nu_1}^{\infty}\frac{4\pi J_{\nu} a_{\nu}(X^i) d\nu}{h\nu},
	\label{ionBalance}
\end{equation}
where $n(X^i)$, $\alpha(X^i)$, $a_{\nu}(X^i)$, $n_e$ and $\nu_1$ are the number density of the $i^{\rm{th}}$ ionization state of species $X$, recombination coefficient, absorption cross section, electron number density and the threshold frequency for ionization of species $X^i$ respectively. Using the MC estimators described in Eqn.~\ref{energydensityMC}, Eqn.~\ref{ionBalance} can then be approximated as
\begin{equation}
	\frac{n(X^{i+1})}{n(X^i)} = \frac{ \epsilon}{\Delta t V \alpha(X^i) n_e} \sum\frac{l a_{\nu}(X^i)}{  h\nu}.
	\label{ionBalanceMC}
\end{equation}

This approach has the advantage that photon energy packets contribute to the estimate of the radiation field without having to undergo absorption events, thus even very optically thin regions are properly sampled. Photoionization calculations are performed iteratively, doubling the number of photon packets per iteration until the temperature and ionization fractions converge.
Charge exchange recombination from O~{\sc ii}, O~{\sc iii} and N~{\sc iv} and ionization of N~{\sc i} and O~{\sc i} is included following the prescription from \citet{King96}

\subsection{Chemistry in photoionized regions}

We include a range of atomic constituents: hydrogen, helium, carbon, nitrogen, oxygen, neon and sulfur, for which we solve the ionization balance using Eqn. \ref{ionBalanceMC}. Note that the ionization balance is solved for every species in the calculation. The ionization states that we treat for metals are C~({\sc i}--{\sc iv}), N~({\sc i}--{\sc iv}), O~({\sc i}--{\sc iii}), Ne~({\sc ii}--{\sc iii}), S~({\sc ii}--{\sc iv}). The hydrogen, helium and C~{\sc iv} recombination rates used by \textsc{torus} are calculated based on \cite{Vern96a}. Other radiative recombination and dielectronic recombination rates are calculated using fits to the results of \cite{Nuss83}, \cite{Pequ91} or \cite{Shul82}. The photoionization cross sections of all atomic species in this paper are calculated using the \textsc{phfit2} routine from \cite{Vern96b}. 

\subsection{Thermal Balance}

\textsc{torus} performs photoionization calculations that incorporate a range of atomic species and in which the thermal balance in each cell is calculated by iterating on the temperature until the heating and cooling rates match. Similarly to the photoionization calculation (Eqn.~\ref{ionBalanceMC}), the heating rate in a given cell is calculated based on the summation of trajectories of photon packets through the cell. This is used to estimate the heating contributions from the photoionization of hydrogen and helium \citep[][]{Wood04} and the heating due to dust photoelectric ejection \citep[][]{Lucy99}. Adding up these three terms give the total heating rate. 

On the other hand, the cooling rate is initially calculated for the maximum and minimum allowed temperatures in the calculation ($3\times10^4\,{\rm K}$ and $10\,{\rm K}$ respectively by default in \textsc{torus}). This is then refined by bisection until the cooling rate matches the heating rate. The cooling processes considered are free--free radiation, hydrogen and helium recombination, dust cooling and collisional excitation of hydrogen and metals.

%%%%%%%%%%%%%%%%%%%%%%%%%%%%%%%%%%%%%%%%%%%%%%%%
\section{coupling {\sc torus} and {\sc 3d-pdr}}%
%%%%%%%%%%%%%%%%%%%%%%%%%%%%%%%%%%%%%%%%%%%%%%%%
\label{sec:coupling}

The strategy we use to couple the two codes is to dismantle {\sc 3d-pdr} and re-assemble it using the {\sc torus} framework\footnote{Our future plans include a publicly available version of {\sc torus-3dpdr}.}. Although the iteration procedure for determining the thermal balance, as well as the associated routines for estimating the heating and cooling functions, are identical to those of the {\sc 3d-pdr} code, we have modified the ray tracing scheme in order to overcome issues related to the large memory requirements. The most important feature however, is the ability to use the MC radiation transport to compute an accurate structure of the UV field in any given arbitrarily complex distribution either for a single or multiple radiation sources. {\sc torus-3dpdr} calculates steady states and its primary use is to post-process snapshots of hydrodynamical simulations. Below we discuss how the code calculates the UV field from the MC photoionization. In Appendix \ref{app:tech} we discuss the technical details followed to couple {\sc torus} and {\sc 3d-pdr}.

\subsection{Calculating the UV field from Monte Carlo photoionization}
\label{ssec:uvMCRT}

In PDR modelling the UV field is usually prescribed at the ionization front and is then assumed to be attenuated according to some simple function (i.e. an exponential). In the coupled code we calculate the UV field using the propagation of photon packets in a photoionization calculation. The UV field in a given cell (in Draines) is the sum of the path lengths, $l$, of photon packets with frequencies in the UV ($912-2400\,$\AA) and is given by 
\begin{equation}
U = \frac{\epsilon}{\Delta t(1.71H_{\rm o})}
        \frac{1}{V}\sum l\hat{u},
\end{equation}
where $\hat{u}$ is the mean radiation propagation vector, $H_{\rm o}=1.6\times10^{-3}\,{\rm erg}\,{\rm s}^{-1}\,{\rm cm}^{-2}$ is the Habing constant \citep{Habi68}, and $1.71$ is the conversion factor between the Habing field to the Draine field \citep{Drai78}. In practice we store the UV field as two vectors, one for positive and one for negative directions. This is important for multi--dimensional calculations, in particular if there are multiple sources (and the radiation field is not simply divergent about a point).

%%%%%%%%%%%%%%%%%%%%%%%%%%
\section{Code evaluation}%
%%%%%%%%%%%%%%%%%%%%%%%%%%
\label{sec:benchmark}

In this section we benchmark {\sc torus-3dpdr} against {\sc 3d-pdr} and {\sc mocassin} for simple one dimensional tests. The PDR abundances correspond to those of Milky Way solar undepleted abundances. The chemical network used here is the most recent UMIST 2012 network \citep{McEl13} in contrast to the \citet{Wood07} version used in B12. The cosmic-ray ionization rate is taken to be $\zeta_{\rm cr}=5\times10^{-17}\,{\rm s}^{-1}$. We also account for the rate of H$_2$ formation on grains using the treatment of \citet{Caza02, Caza04} and we determine the dust temperature following the treatment of \citet{Holl91} in addition to the {\sc 3d-pdr} version presented in B12.

\subsection{Testing the UV intensity}
\begin{figure}
	\includegraphics[width=0.5\textwidth]{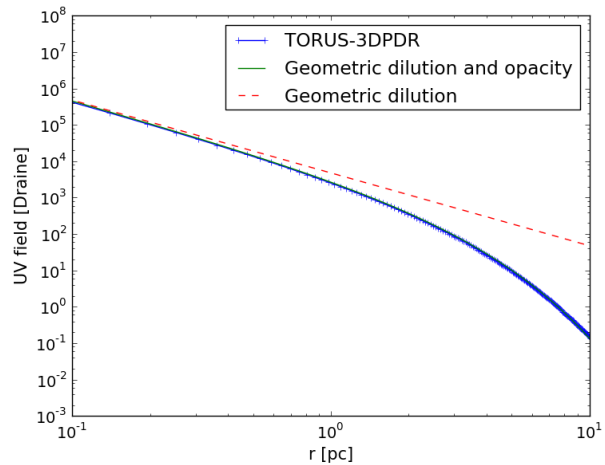}
	\caption{ This plot shows the extinction of the UV field in {\sc torus-3dpdr} (blue crossed line) versus Eqn. \ref{eqn:uvtest} with the opacity term $e^{-\kappa r}$ (green solid line) and without this term (red dashed line). {\sc torus-3dpdr} reproduces precisely Eqn. \ref{eqn:uvtest} verifying that the UV estimate is accurate.}
	\label{fig:uvtest}
\end{figure}

To check whether the UV estimator we use is accurate, we calculate the UV field as a function of distance from a test source. The source that we consider is the same as that used in the HII40 Lexington benchmark. We compare with our numerical estimate by integrating over the UV band ($912 - 2400$ \AA) of the stellar spectrum and attenuate this value accounting for both the distance from the source and the gas opacity, i.e.
\begin{eqnarray}
\label{eqn:uvtest}
\chi=\frac{\chi_{\circ}}{4\pi r^2}e^{-\kappa r}
\end{eqnarray}
where $\chi$ is the attenuated $\chi_{\circ}$ UV field, $r$ is the distance from the source, and $\kappa$ is the opacity term (see Appendix \ref{app:dilution}). We have compared the UV distribution (by invoking the on-the-spot approximation) calculated by the {\sc torus-3dpdr} with the above Eqn. \ref{eqn:uvtest} and we show the results in Fig. \ref{fig:uvtest}. It can be seen that the agreement is excellent, verifying that our UV estimate is accurate.

\subsection{Consistency between {\sc 3d-pdr} and {\sc torus-3dpdr}}
In order to assess the accuracy of the PDR calculations after the coupling of the two codes, we perform a one-dimensional test to compare the coupled {\sc torus-3dpdr} code against {\sc 3d-pdr}. The setup of the test we adopt is very similar to the V2 model discussed in the \citet[][ hereafter `R07']{Roll07} paper. However, due to the recent updates of the {\sc 3d-pdr} code described above, we do not aim at benchmarking {\sc torus-3dpdr} against the various other one-dimensional PDR codes described in R07. The adopted model consists of a one-dimensional uniform density distribution with H-nucleus number density of $n_{\rm H}=10^5\,{\rm cm}^{-3}$. The size of the distribution is chosen so that the maximum visual extinction is $A_{\rm V,max}=10\,{\rm mag}$. We assume that it is irradiated by a plane-parallel radiation field of strength $\chi_0=10^3$ multiples of the Draine radiation field \citep{Drai78}, and that the attenuation of the UV field is according to Eqn. \ref{eqn:3dpdrUV}.

As described in B12, in order to emulate a semi-infinite slab we have considered cells aligned along two opposite {\sc healpix} rays at the $\ell=0$ level of refinement while assuming very high optical depths in all other directions. Here, the {\sc 3d-pdr} run uses ${\cal N}_{Av}=20$ elements logarithmically distributed per $A_{\rm V}$ dex with $-5\le\log(A_{\rm V})\le1$ in all cases, implying a total number of $N_{\rm elem}=120$ elements. The {\sc torus-3dpdr} runs use a non-uniform mesh, clustered near the ionization front. The UV field is assumed to be plane-parallel impinging from one side while the attenuation of the field is calculated using Eqn. \ref{eqn:3dpdrUV}. 

\begin{figure*}
	\includegraphics[width=8cm]{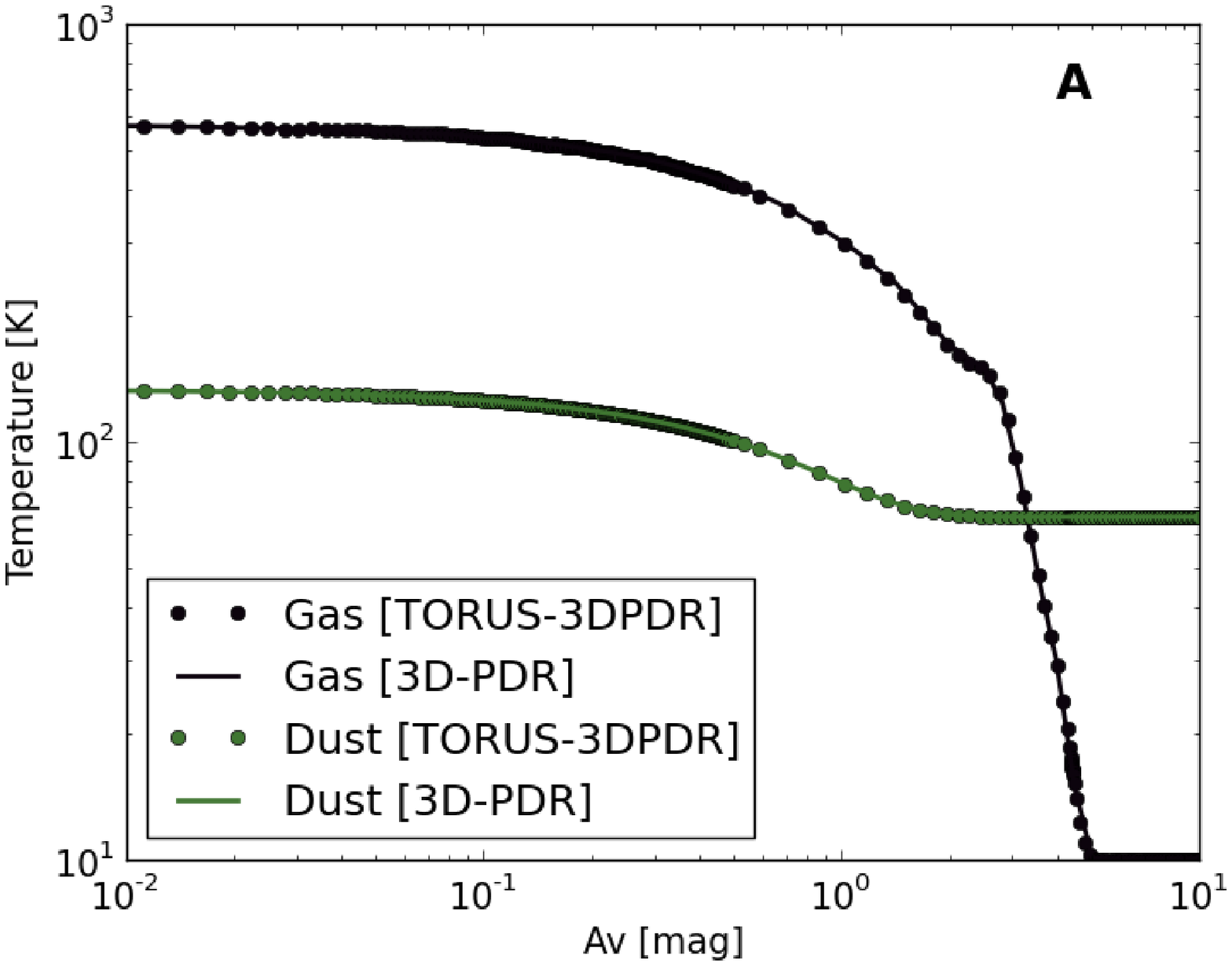}
	\includegraphics[width=8cm]{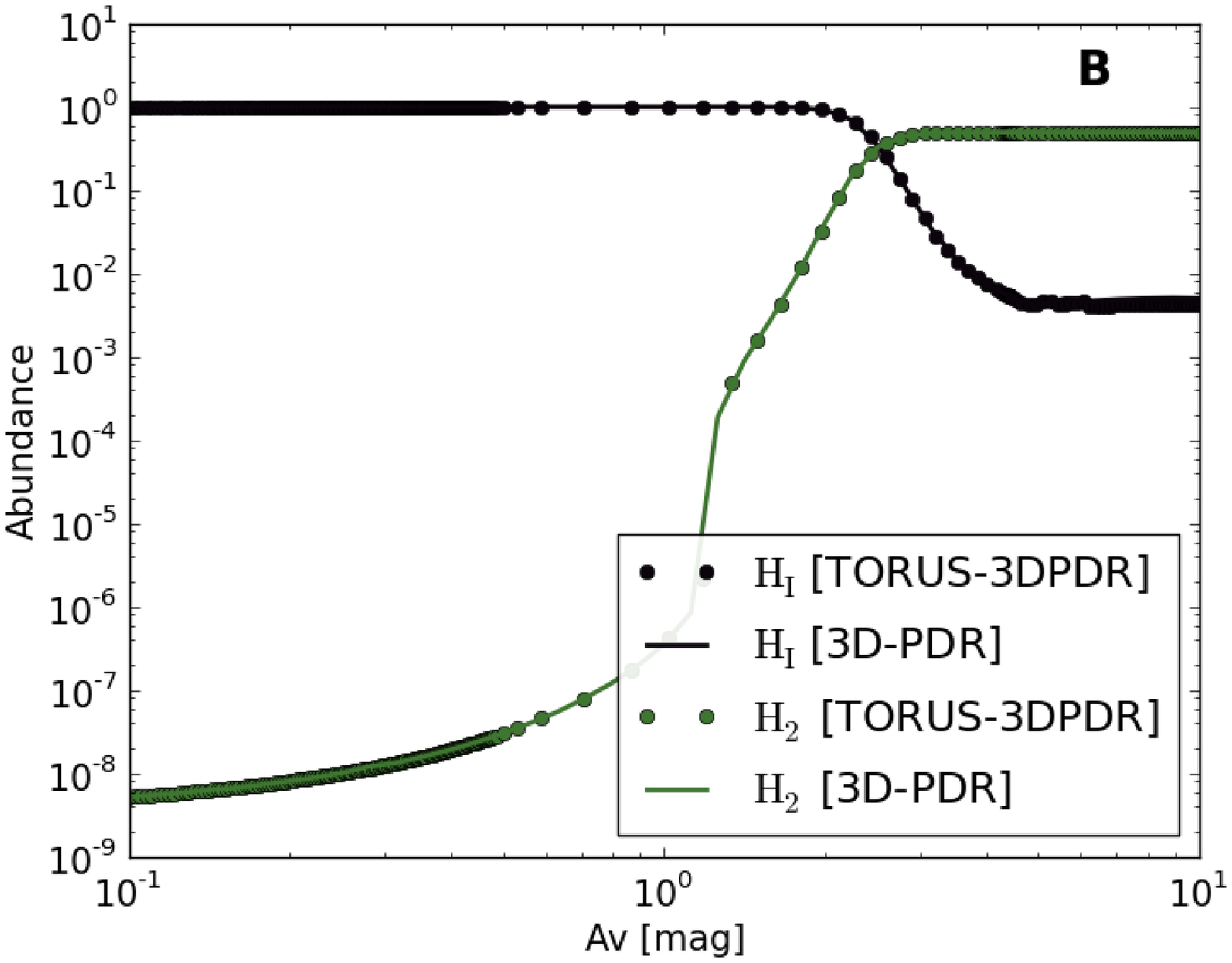}
	\includegraphics[width=8cm]{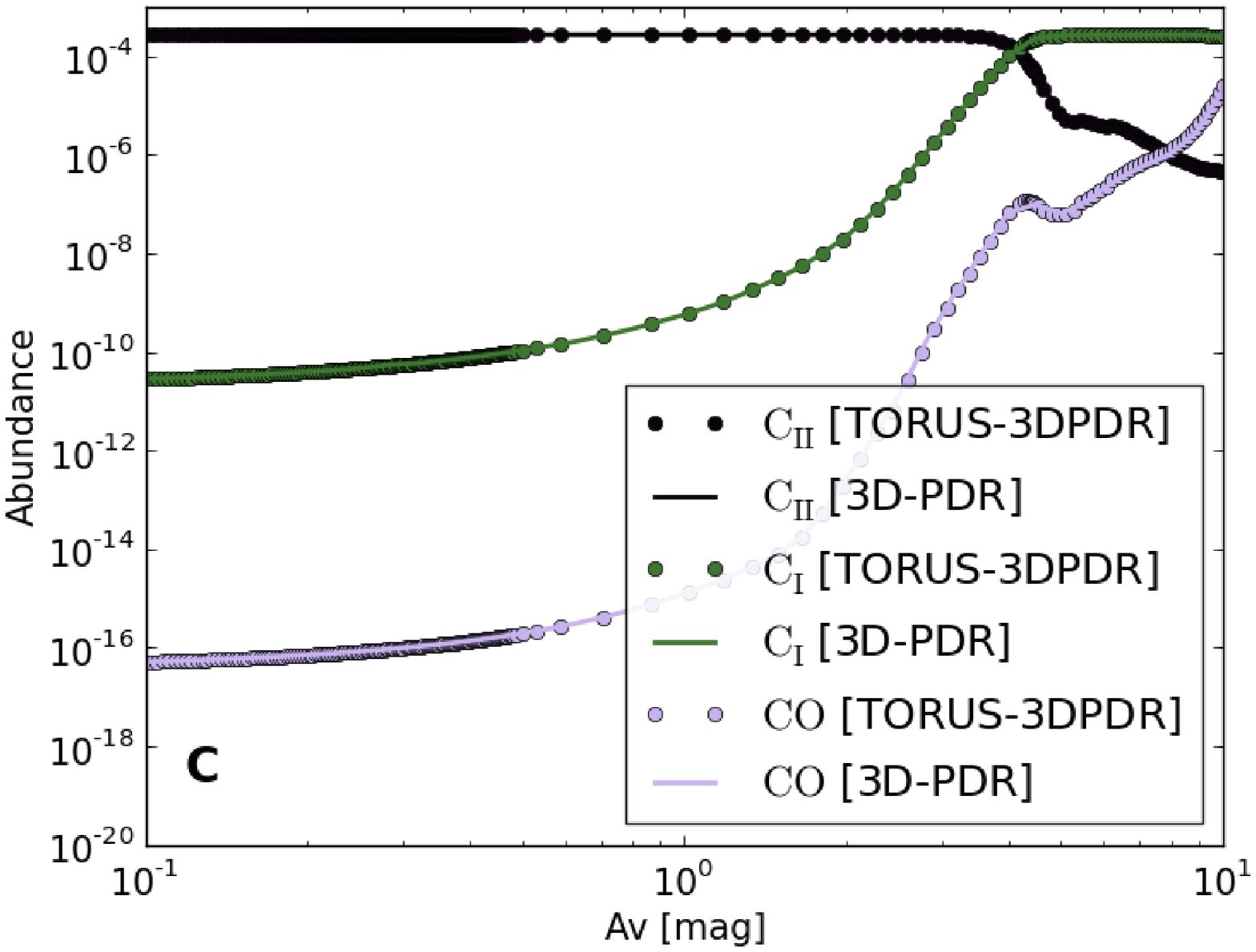}
	\includegraphics[width=8cm]{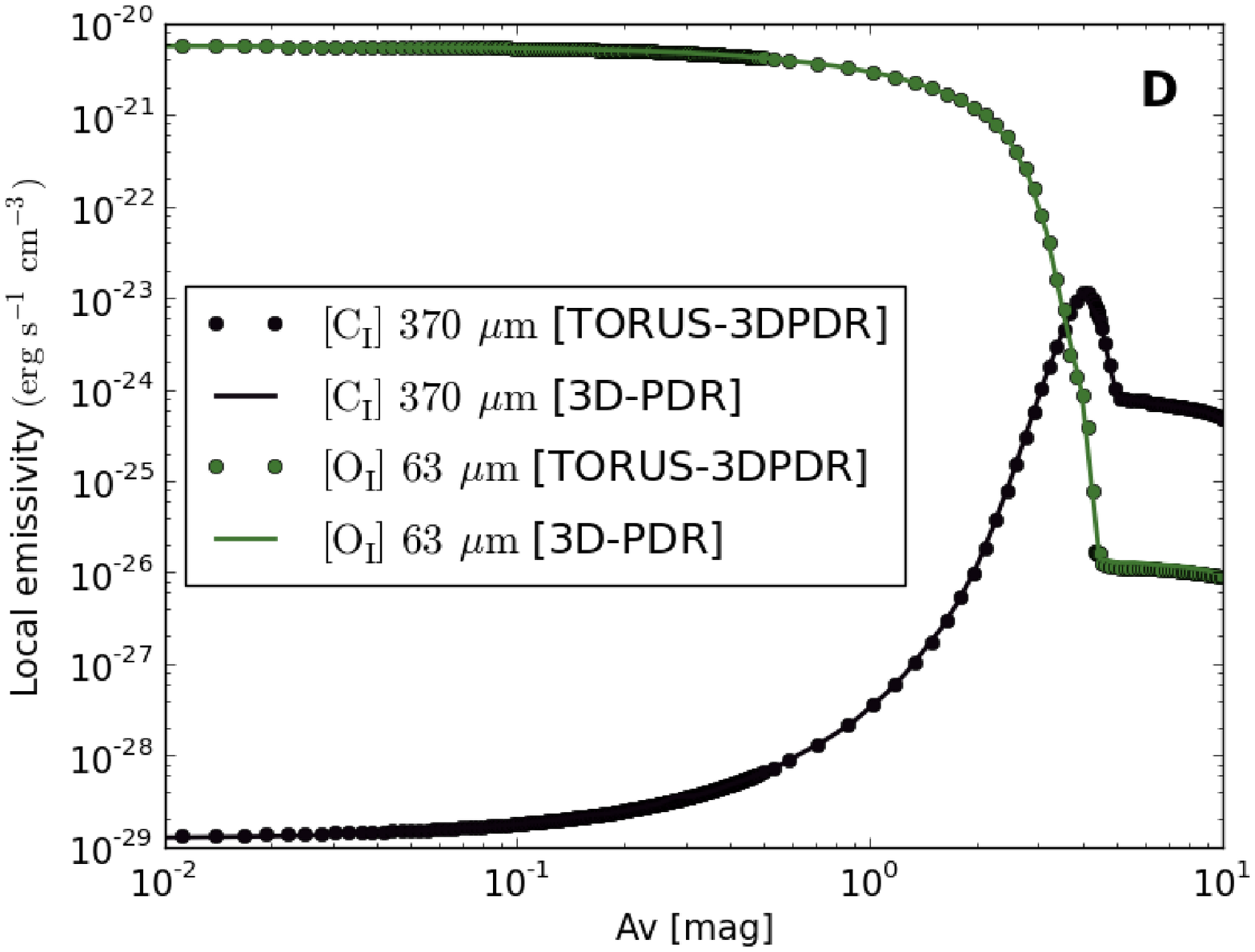}
	\includegraphics[width=8cm]{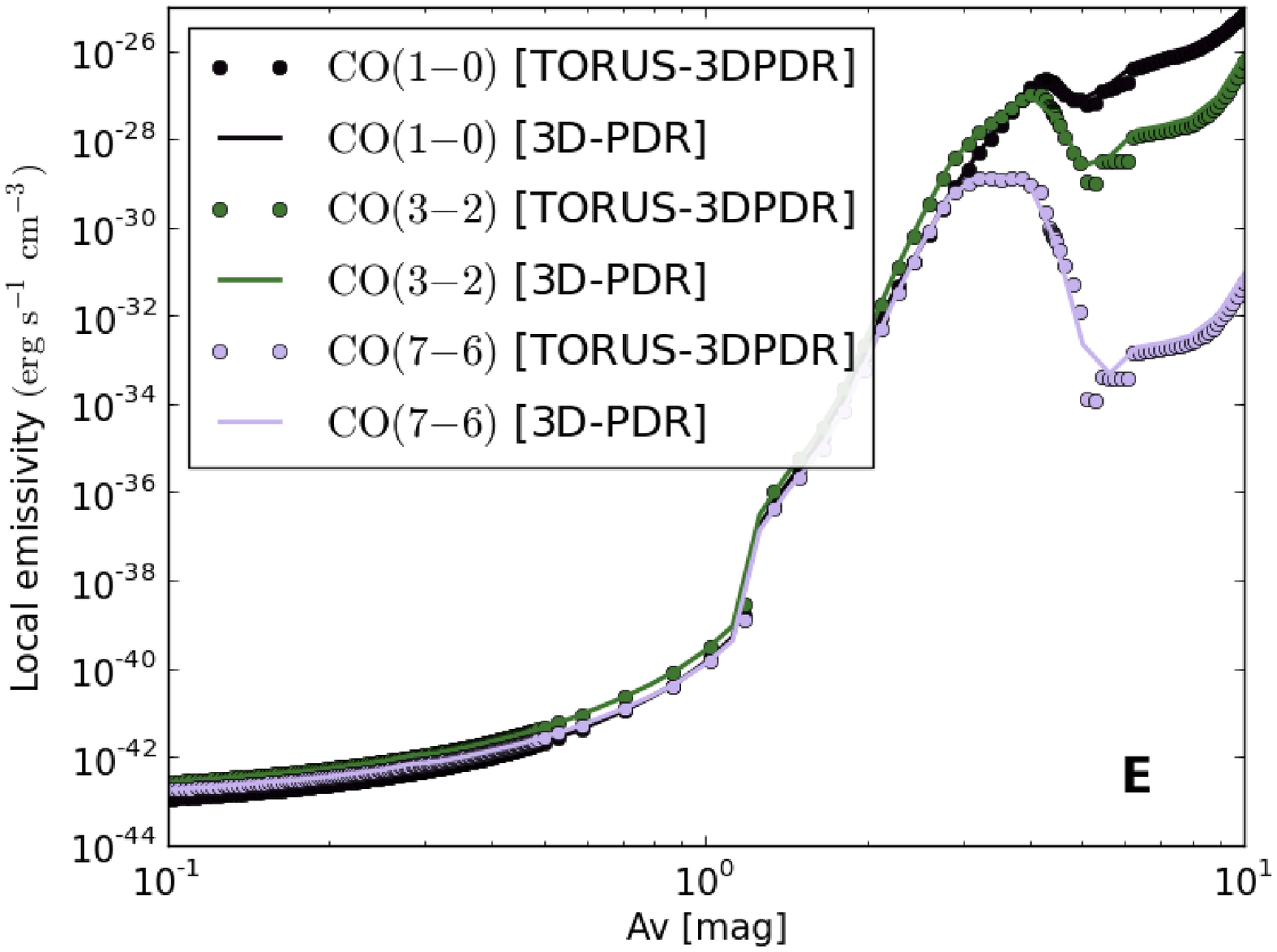}
	\includegraphics[width=8cm]{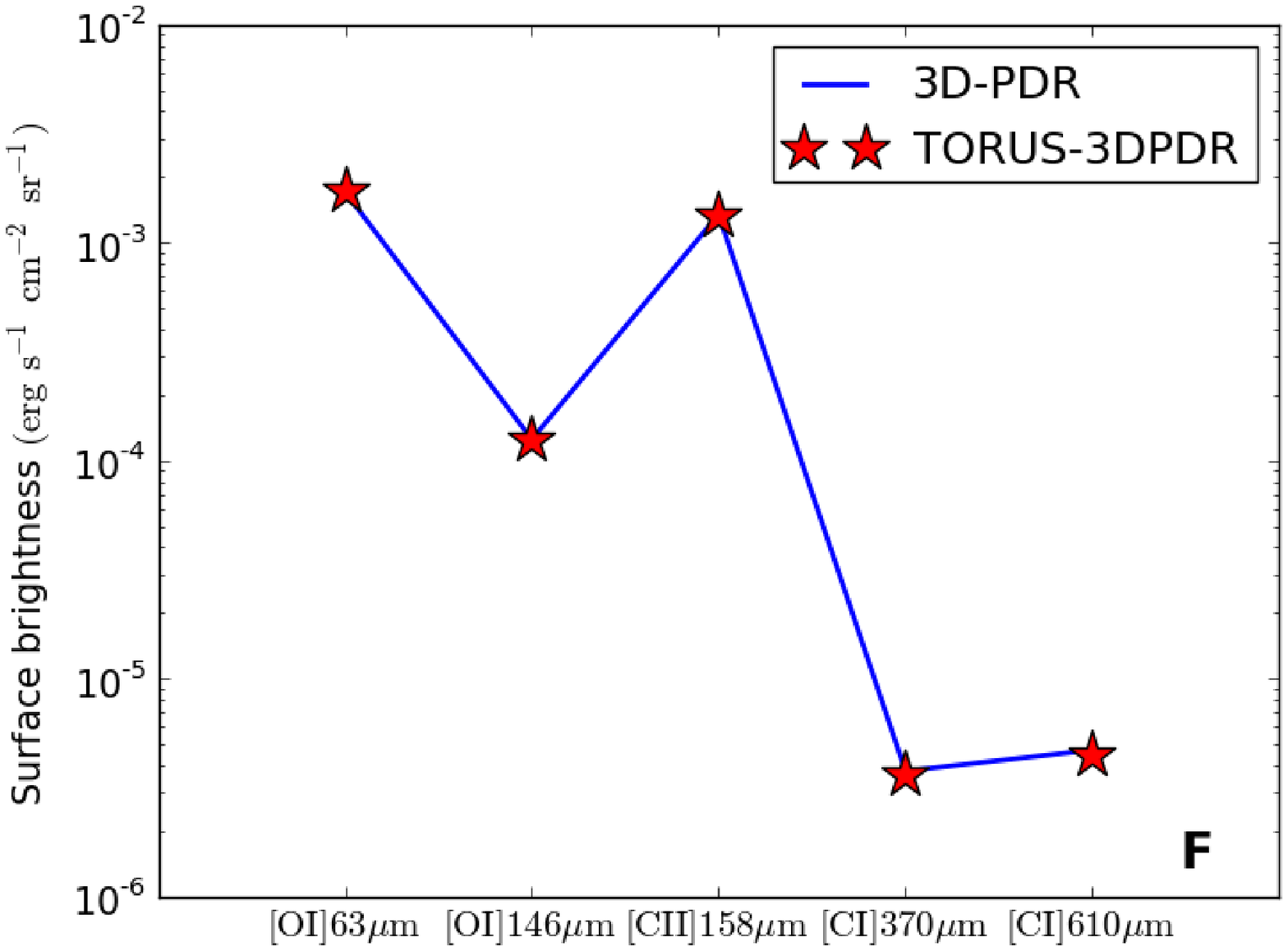}
	\caption{Comparison between {\sc 3d-pdr} (solid lines) and {\sc torus-3dpdr} (filled circles) for the benchmark model V2. Panel (a) shows the gas and dust temperature profiles. Panel (b) shows the H and H$_2$ abundances while panel (c) shows the C~{\sc ii}, C~{\sc i} and CO abundances versus the visual extinction $A_{\rm V}$. Panel (d) shows the local emissivities for [C~{\sc i}]~370$\mu$m and [O~{\sc i}]~63$\mu$m and panel (e) shows the local emissivities for three different CO line transitions. Finally, panel (f) shows the surface brightnesses for the dominant cooling lines. In all cases the agreement is excellent. The discrepancy observed in the bottom left panel at $A_{\rm V}\sim5$ is due to noise resulting from the grid.}
	\label{fig:Roll1}
\end{figure*}

Figure \ref{fig:Roll1} shows results for the benchmark model described above. We compare gas and dust temperature profiles, abundances of H~{\sc i}, H$_2$, C~{\sc ii}, C~{\sc i} and CO versus $A_{\rm V}$, and local emissivities as well as emergent intensities (surface brightnesses) for some of the dominant cooling lines. The agreement between the two codes is excellent. We also note that we have performed all other tests described in the R07 work (i.e. V1, V3, and V4) and we have found excellent agreement as well.

\subsection{Photoionization Benchmarking}
The HII40 Lexington benchmark \citep{Ferl95,Pequ01} is a canonical test of photoionization codes. It involves modelling the ionization and temperature structure of a uniform density medium around a massive star. It is usually performed in 1D owing to the spherically symmetric nature of the problem. \cite{Hawo12} demonstrated the veracity of the photoionization scheme in \textsc{torus} using this test, obtaining agreement with the ionization and temperature structure computed by the well known \textsc{cloudy} code \citep{Ferl98}.

Table \ref{tab:lex} compares {\sc torus} with the median values of all participating codes in the HII40 Lexington benchmark \citep{Pequ01}. Overall we find good agreement between the lines emmisivities. 

\begin{table}
 \centering
  \caption{Comparison between {\sc torus} with the median value for all participating codes in the Lexington 2000 benchmark workshop for the standard H~{\sc ii} region (HII40) test. The values of the `median' column are those listed in Table 4 of \citet{Erco03}. The third column corresponds to the absolute error $100\cdot|I_{\rm TORUS}-I_{\rm med}|/|I_{\rm med}|$, where $I_{\rm med}$ is the median value and the $I_{\rm TORUS}$ is the {\sc TORUS} result for the intensity of each line.}
  \label{tab:lex}

  \begin{tabular}{l c c c}
  \hline
   Line & Median & {\sc torus} & $\sim$Error\%  \\
     \hline
   H$\beta/10^{37}\,{\rm erg}\,{\rm s}^{-1}$   & 2.05     & 1.91    &  7    \\
   C~{\sc iii}$]$ $1907+1909$		       & 0.070    & 0.142   &  51    \\
   $[$N~{\sc ii}$]$ $122\,\mu{\rm m}$          & 0.034    & 0.027   &  26    \\
   $[$N~{\sc ii}$]$ $6584+6548$                & 0.730    & 0.801   &  9     \\
   $[$N~{\sc ii}$]$ $5755$		       & 0.0054   & 0.0077  &  30     \\
   $[$N~{\sc iii}$]$ $57.3\,\mu{\rm m}$        & 0.292    & 0.263   &  11    \\
   $[$O~{\sc i}$]$ $6300+6363$                 & 0.0086   & 0.0276  &  69     \\
   $[$O~{\sc ii}$]$ $7320+7330$                & 0.029    & 0.022   &  32    \\
   $[$O~{\sc ii}$]$ $3726+3729$                & 2.03     & 2.57    &  21   \\
   $[$O~{\sc iii}$]$ $5007+4959$               & 2.18     & 2.80    &  22   \\
   $[$O~{\sc iii}$]$ $4363$                    & 0.0037   & 0.0108  &  66     \\
   $[$O~{\sc iii}$]$ $52+88\,\mu{\rm m}$       & 2.28     & 2.77    &  18   \\
   $[$Ne~{\sc ii}$]$ $12.8\,\mu{\rm m}$        & 0.195    & 0.171   &  14    \\
   $[$Ne~{\sc iii}$]$ $15.5\,\mu{\rm m}$       & 0.322    & 0.350   &  8     \\
   $[$Ne~{\sc iii}$]$ $3869+3968$              & 0.085    & 0.110   &  23    \\
   $[$S~{\sc ii}$]$ $6716+6731$                & 0.147    & 0.159   &  8     \\
   $[$S~{\sc ii}$]$ $4068+4076$                & 0.0080   & 0.0062  &  29     \\
   $[$S~{\sc iii}$]$ $18.7\,\mu{\rm m}$        & 0.577    & 0.741   &  22    \\
   $[$S~{\sc iii}$]$ $9532+9069$               & 1.22     & 1.23    &  $<$1   \\

 \hline
\hline
\end{tabular}
\end{table}

\subsection{Photoionization + PDR test model}
\label{ssec:photopdr}

In this section we perform a one-dimensional test application in which the integrated scheme is being tested to simulate the interaction of a uniform-density cloud with ionizing radiation. The density, $n_{\rm H}$, is taken to be $100\,{\rm cm}^{-3}$ and the star is assumed to have a blackbody temperature of $4\times10^4\,{\rm K}$ and radius $18.67{\rm R}_{\odot}$ emitting photons from the left-hand edge of the grid. The grid is taken to be uniform and spherical consisting of 256 cells. These values correspond to the HII40 simulation of the Lexington benchmarking \citep{Ferl95}.

We convert the mean intensity into Draine units using the relation:
\begin{eqnarray}
\label{eqn:UVtoDra}
\chi\,=\frac{\int_{912}^{2400}J_{\lambda}d\lambda}{1.71H_{\rm o}}\,
\end{eqnarray}
where $J_{\lambda}$ is the intensity of the radiation integrated over wavelength (here in \AA). Assuming monochromatic radiation at $\lambda=912\AA$, the integral of the above equation can be considered as a $\delta$-function and Eqn. \ref{eqn:UVtoDra} can therefore take the form:
\begin{eqnarray}
\chi\,=\frac{\dot{\cal N}_{\rm LyC}}{4\pi r^2}\frac{hc}{912\AA}\frac{1}{1.71H_{\rm o}}.
\end{eqnarray}
The above equation calculates the intensity of the UV radiation field (in ${\rm Draine}$ units) resulting from a star emitting $\dot{\cal N}_{\rm LyC}$ photons per unit time placed at distance $r$.

\begin{figure}
	\includegraphics[width=0.5\textwidth]{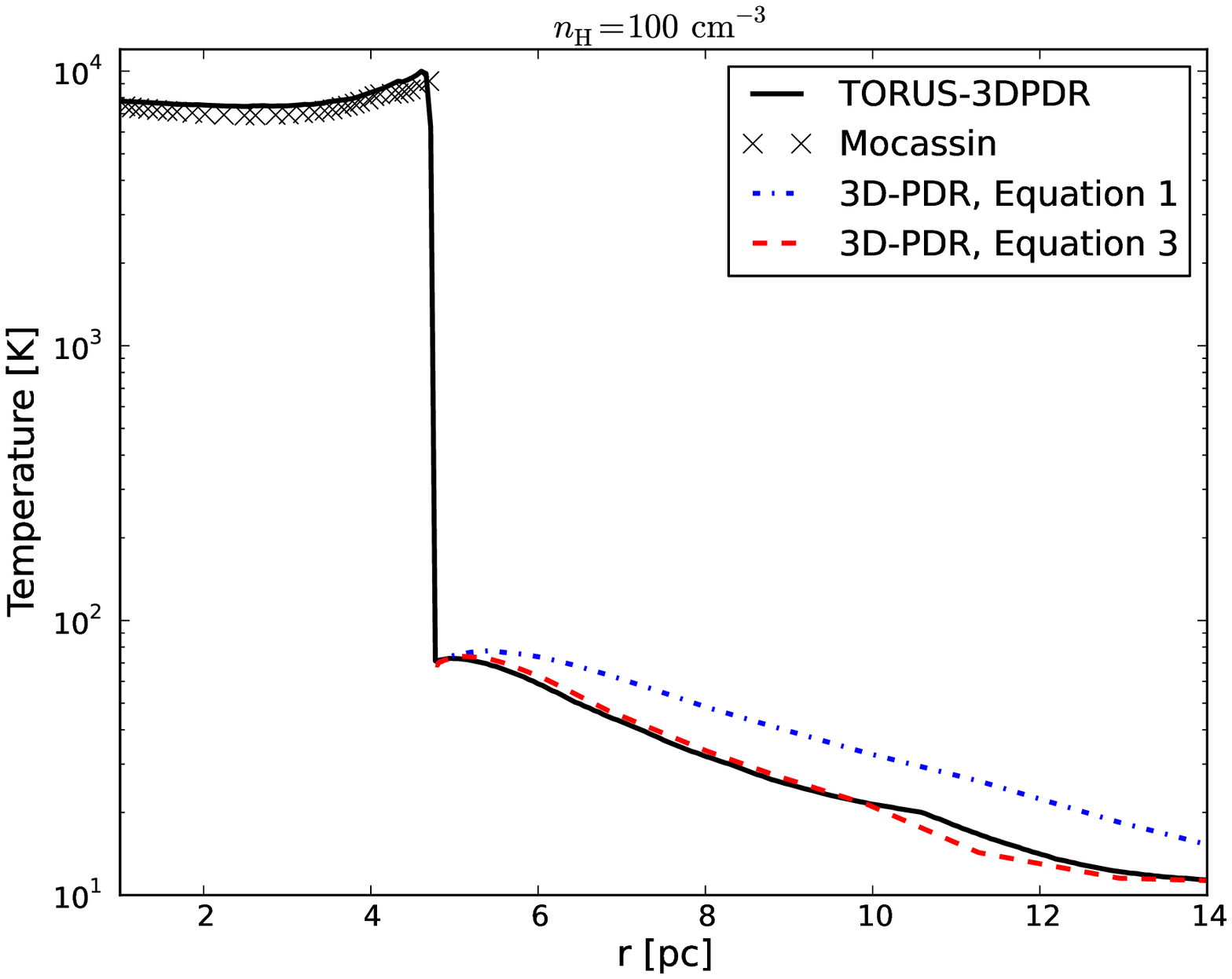}
	\includegraphics[width=0.5\textwidth]{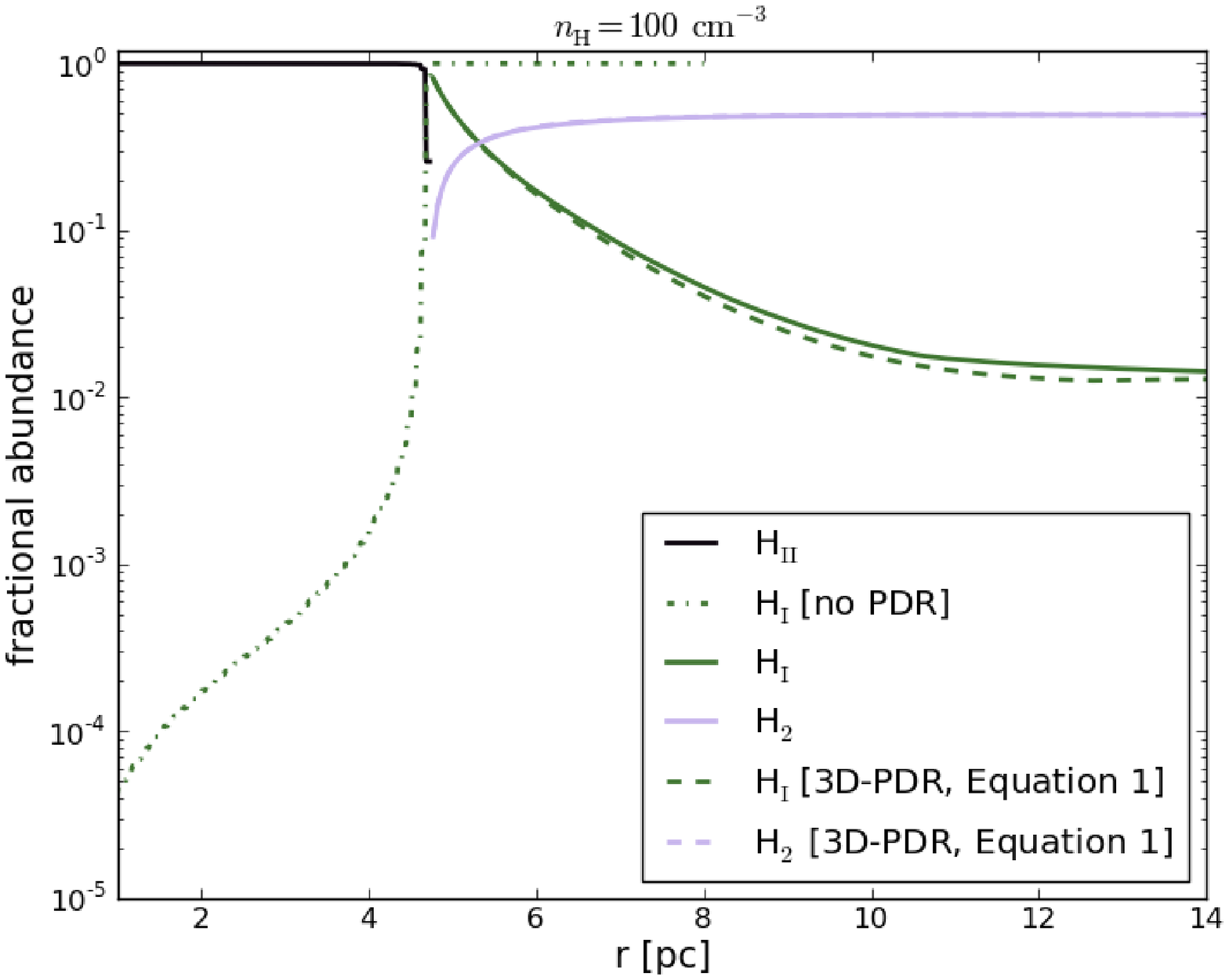}
	\caption{ Upper panel: Gas temperatures obtained by {\sc torus-3dpdr} (black solid line) and {\sc 3d-pdr} (red dashed, and blue dot-dashed). The blue dot-dashed line corresponds to the attenuation of the UV field as described by Eqn. \ref{eqn:3dpdrUV}, whereas the red dashed line is as described by Eqn. \ref{eqn:geomdil}. Black crosses (+) show a comparison with the {\sc mocassin} code. Lower panel: fractional abundances of hydrogen for different phases. The black solid line is ionized H~{\sc ii}, the green solid line is atomic H~{\sc i}, and the blue-gray solid line is molecular H$_2$. The green dot-dashed line is atomic hydrogen obtained without PDR calculations (i.e. as in the Lexington test). All these correspond to the {\sc torus-3dpdr} code. The green and blue dashed lines correspond to {\sc 3d-pdr} runs using Eqn. \ref{eqn:3dpdrUV} to account for the attenuation of the radiation field.}
	\label{fig:fullrun}
\end{figure}

We perform two runs in which we explore the effect of the attenuation of radiation in the PDR scheme estimated i) directly from the MC approach as calculated in {\sc torus} and described in \S\ref{ssec:uvMCRT} and ii) using the exponential relation given by Eqn. \ref{eqn:3dpdrUV}. {\sc torus-3dpdr} selects as PDR cells any cell that has a temperature below 3000K corresponding to an ionization fraction of $\chi_{\rm i}\le0.3$. The upper panel of Fig. \ref{fig:fullrun} shows the temperature profiles while the lower panel shows the fractional abundances of H~{\sc ii}, H~{\sc i}, and H$_2$. The ionizing radiation is impinging from the left hand side in the above diagrams (the position of the star defines the centre of the Cartesian co-ordinate system). In addition, in the upper panel we overplot the results obtained from {\sc mocassin} \citep{Erco03} for the calculations concerning the photoionized region, while we also overplot those obtained from {\sc 3d-pdr} for the photodissociation region. We find that the integrated code {\sc torus-3dpdr} is in very good agreement with {\sc mocassin}. For the particular case of the MC-attenuation, we have performed an additional {\sc 3d-pdr} calculation in which we have included the additional extinction due to the distance of the point source from each PDR cell as described by Eqn. \ref{eqn:geomdil}. As is shown in the upper panel of Fig. \ref{fig:fullrun}, the PDR temperature profile obtained from the MC {\sc torus-3dpdr} technique is reproducible only when the distance from the source is taken into account. Note, however, that H$_2$ and H~{\sc i} are not strongly sensitive to that.

%%%%%%%%%%%%%%%%%%%%%%%
\section{Applications}%
%%%%%%%%%%%%%%%%%%%%%%%
\label{sec:applications}

In this section we present the capabilities of the code in simulating three-dimensional density structures interacting with one- or two- sources, either by invoking the on-the-spot approximation or including the diffusive component of the ionizing radiation, thus exploring the importance of these components in regulating the chemistry of PDRs. To do this we consider an oblate spheroidal cloud with a uniform density distribution. The choice of such an object was based on the assumption it is not spherically symmetric hence not reproducible by an equivalent one-dimensional approach, while at the same time it offers simplified results. 

\begin{figure}
  \includegraphics[width=8cm]{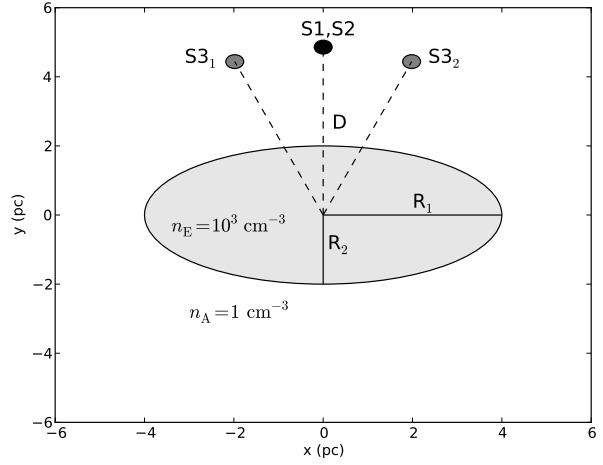}
  \caption{ Setup of the three-dimensional oblate spheroidal cloud case discussed in \S\ref{sec:applications}. The $x-$ and $y-$ axis dimensions are in ${\rm pc}$. The shaded spheroid has dimensions of $R_1=4\,{\rm pc}$ and $R_2=2\,{\rm pc}$ and it has a uniform density with $n_{\rm E}=10^3\,{\rm cm}^{-3}$. The center of the cloud defines the center of a Cartesian co-ordinate system. The spheroid is embedded in an ambient medium with density $n_{\rm A}=1\,{\rm cm}^{-3}$. The middle black point corresponds to the position of the ionizing source for the simulations, S1 (\S\ref{ssec:S1}) and S2 (\S\ref{ssec:S2}). The gray points, S3$_1$ and S3$_2$, correspond to the two positions of the stars in the S3 simulation (\S\ref{ssec:S3}). All stars are placed at a distance $D$ from the center of the oblate spheroidal cloud.}
  \label{fig:simsetup}
\end{figure}

We consider a cubic computational domain with size $R=12\,{\rm pc}$ consisting of $64^3$ cells of equal volume. Figure \ref{fig:simsetup} shows the simulation setup. The center of the spheroid defines the center of a Cartesian co-ordinate system. The semi-major axis is taken to be $R_1=4\,{\rm pc}$, the semi-minor axis is taken to be $R_2=2\,{\rm pc}$ and the density is assumed to be $n_{\rm E}=1000\,{\rm cm}^{-3}$. This means that the visual extinction, $A_{\rm V}$, can reach values up to $\sim8\,{\rm mag}$ along the $y-$axis. We can consider this structure as a cloud of mass $M_{\rm E}=1655\,{\rm M}_{\odot}$. The spheroid is embedded in an ambient medium of uniform density $n_{\rm A}=1\,{\rm cm}^{-3}$. In all simulations we use the MC photoionization routine described in \S\ref{ssec:mcphot} and we use an $\ell=0$ {\sc healpix} level of refinement as explained in \S\ref{ssec:hlevel}.

We perform three simulations. In the first simulation (S1), we place a single exciting source at a distance of $D=4.86\,{\rm pc}$ from the center of the spheroid emitting $\dot{\cal N}_{\rm LyC}=1.2\times10^{50}$ photons per second and invoking the on-the-spot approximation. In the second simulation (S2), we use the same setup but we switch on the contribution of the diffuse component of the radiation field. Thus by comparing simulations S1 and S2 we examine the effect of the diffuse radiation field on the abundance distribution. In the third simulation (S3) we replace the single exciting source with two sources emitting half the number of photons per second, compared to the S1 and S2 cases. In particular, we place the two stars at positions $(x,y)=(\pm1.98,4.44)\,{\rm pc}$ which ensures that the distance $D$ from the center of the spheroid is kept the same as in S1 and S2. Each of the two stars emit $\dot{\cal N}'_{\rm LyC}=6\times10^{49}$ photons per second and the diffuse component of radiation is taken into account. Therefore, in the S3 simulation we test the capabilities of the code in treating PDRs interacting with multiple sources and we examine the consequent effects introduced in the abundances distribution as well as on the distribution of the local emissivities of the most important cooling lines. 

The stellar spectra in the models presented here are blackbodies, which are probabilistically randomly sampled by the MC photon packets. There are therefore occasionally high frequency ionizing photon packets, that can propagate far into the neutral gas since the photoionization cross section is proportional to the inverse cube of the photon frequency, i.e. $\propto(\nu_{\rm trsh}/\nu)^3$
%\begin{eqnarray}
%\sigma \propto (\nu_{\rm trsh}/\nu)^3,
%\end{eqnarray}
where the $\nu_{\rm trsh}$ is the ionization threshold (e.g. $13.6\,{\rm eV}/h$ for hydrogen). These high frequency packets are responsible for regions of localized heating exterior to the main ionization front. In the case of our three-dimensional models, the calculations with a single star consider a source that has $R=18\,R_{\odot}$ radius and $T_{\rm eff}=5\times10^4\,{\rm K}$ effective temperature (whereas in the multiple sources simulation each source has $R=12.7\,R_{\odot}$ with half that effective temperature), and so have a higher probability of emitting high frequency photon packets. This is why the temperature distribution within the ellipse and particularly in the ionization front is subject to spurious heating which has a significant (but very local) effect in determining the abundances distribution. These areas can be seen in almost all panels of Fig. \ref{fig:diff1star} and Fig. \ref{fig:difflines} and are located close to the $x=0\,{\rm pc}$ axis (i.e. $x\sim-1\,{\rm pc}$ and $x\sim1\,{\rm pc}$) and with $y>1\,{\rm pc}$.

This can be alleviated by using more photon packets (so the energy per packet is reduced) however at the cost of computational expense. We will thus exclude this effect from all discussion below. We also note that in the particular case of simulation S3 and because the ionizing photons have lower frequency than in the S1 and S2 cases, this effect is not present.

\subsection{Angular resolution in the PDR regime}
\label{ssec:hlevel}

\begin{figure*}
\includegraphics[width=0.95\textwidth]{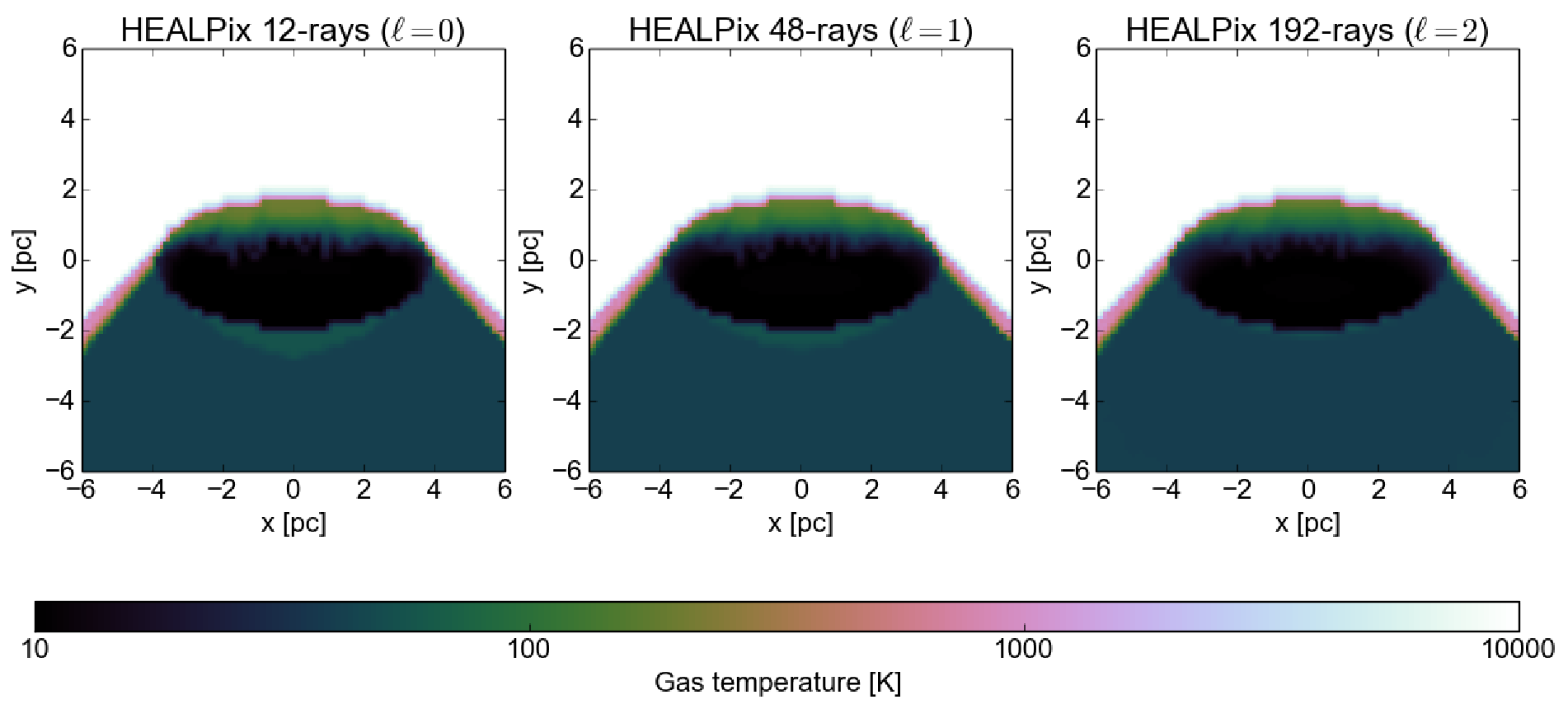}
\includegraphics[width=0.95\textwidth]{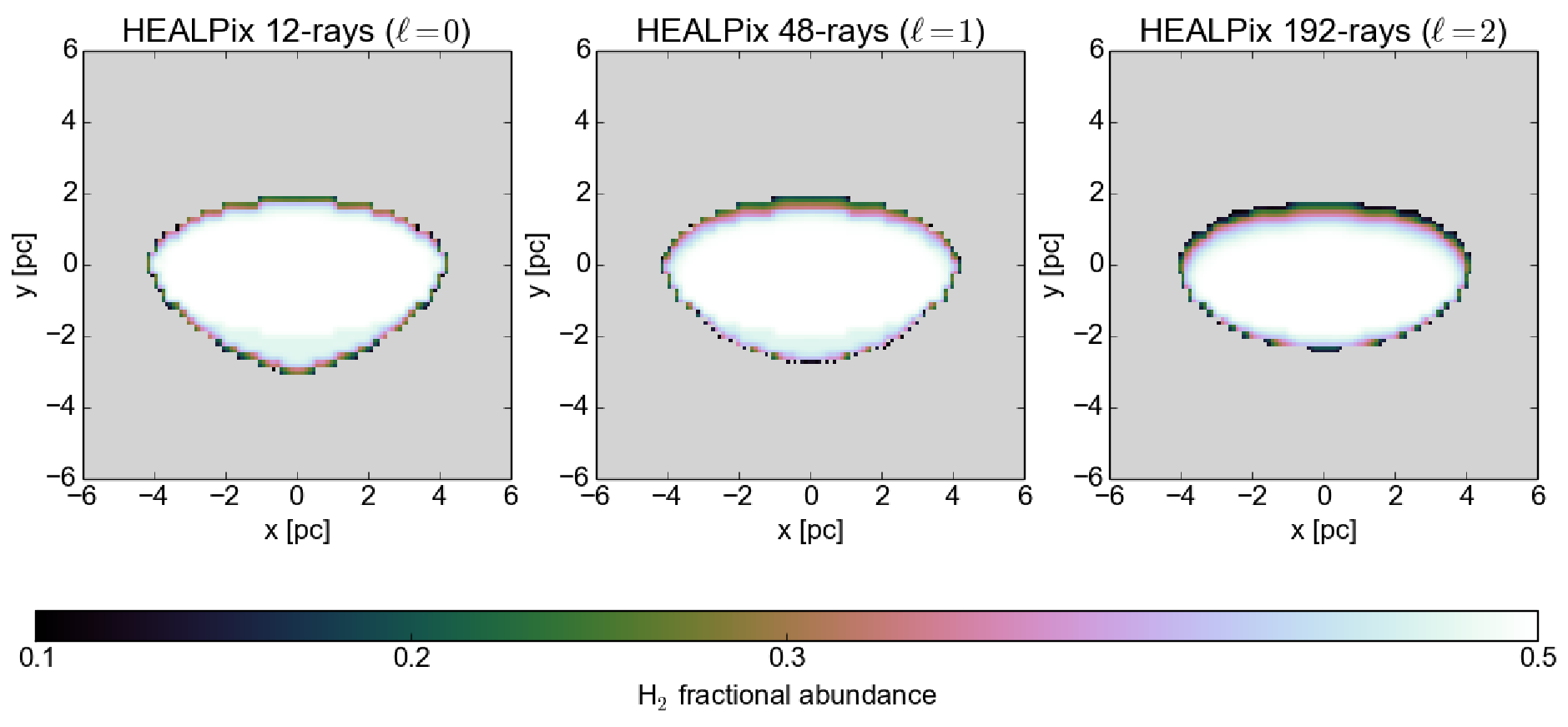}
\caption{ Cross section plots of the S1 simulation using different levels of {\sc healpix} refinement. Top row is gas temperature and bottom row is the abundance distribution of H$_2$. From left to right $\ell=0$ (12-rays), $\ell=1$ (48-rays), $\ell=2$ (192-rays) of refinement. It can be seen that there are no significant differences for the tests presented here when moving to high angular resolution. For all simulations presented in this paper we shall use $\ell=0$. The gray color in the background of the bottom row indicates a zero value.}
\label{fig:healpixtest}
\end{figure*} 

As a first test we explore the effect of using different levels of {\sc healpix} refinement corresponding to different angular resolutions per direction in each cell of the computational domain. We perform two additional simulations to S1, one with $\ell=0$ (corresponding to 12 {\sc healpix} rays) and one with $\ell=2$ (corresponding to 192 {\sc healpix} rays). Figure \ref{fig:healpixtest} shows the changes observed for the gas temperature and H$_2$ abundance distribution. Overall we find excellent agreement in the gas temperature distribution and minor changes in the H$_2$ abundance distribution, showing that even at low angular resolution (i.e. at $\ell=0$) our results remain consistent with those at higher angular resolution (i.e. $\ell=2$). Given also that the computational expense increases approximately 4 times by increasing the {\sc healpix} level (since there are 4 times more rays emanated from each cell; see Appendix \ref{app:cputime} for further discussion) in this paper we will use $\ell=0$ for all simulations. We note that in more complicated simulations with non-uniform density distribution, a higher level of angular resolution must be used.

\subsection{Single source and on-the-spot approximation (S1)}
\label{ssec:S1}

\begin{figure*}
  \includegraphics[width=0.95\textwidth]{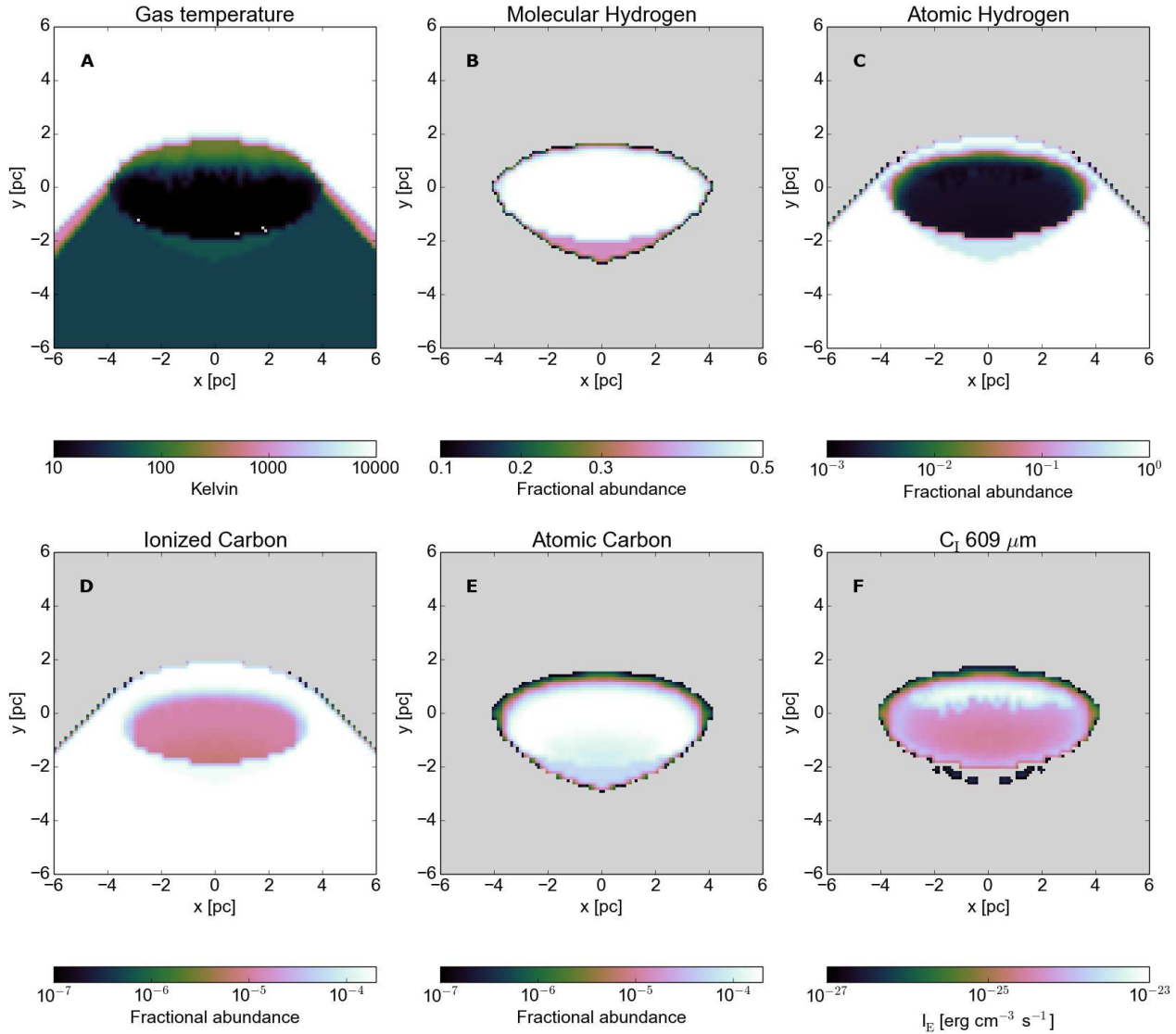}%CrossSectionsOTS.eps} 16cm
  \caption{ Cross section plots for the S1 simulation. The axes are in ${\rm pc}$. Panel A shows the gas temperature distribution. Panels B-E show the distribution of the fractional abundances of molecular hydrogen (H$_2$), atomic hydrogen (H~{\sc i}), ionized carbon (C~{\sc ii}), and atomic carbon (C~{\sc i}) respectively. The abundances of these species are with respect to the total hydrogen nucleus abundance. Panel F shows the local emissivity of [C~{\sc i}]~$609\mu m$. The colour bar in panel F is in units of ${\rm erg}\,{\rm cm}^{-3}\,{\rm s}^{-1}$. In panels B-F the gray background indicates a zero value.}
  \label{fig:1star}
\end{figure*}

Figure \ref{fig:1star} shows the outcomes of the S1 simulation. We show cross-section plots (at $z=0\,{\rm pc}$ plane) of the gas temperature (Fig. \ref{fig:1star}A), the abundance distributions of H$_2$, H~{\sc i}, C~{\sc ii} and C~{\sc i} (Fig. \ref{fig:1star}B-E respectively) and the local emissivity of [C~{\sc i}] $609\,\mu{\rm m}$ (Fig. \ref{fig:1star}F). The ionizing source interacts with the rarefied medium resulting in an H~{\sc ii} region. The dense oblate spheroidal cloud on the other hand shields the propagation of the UV photons behind it. The resultant temperature in the H~{\sc ii} region is $\sim10^4\,{\rm K}$ while in the shadowed region it is $\sim50\,{\rm K}$, and in the innermost part of the spheroid $\sim10\,{\rm K}$. This extra heating, that shifts the temperature from $10\,{\rm K}$ to $\sim40\,{\rm K}$ in the ambient medium behind the spheroid, is due to the interaction of that rarefied gas with cosmic rays, and which do not attenuate inside this computational domain. Note that the gas temperature and ionization fraction transition from the ionized medium to the PDR is very abrupt.

The shielding of the UV radiation allows the formation of atoms and molecules in the interior of the cloud. In the modelled cloud, this shielding extinguishes the UV radiation to such an extent that is able to photodissociate H$_2$ to form H~{\sc i} while it cannot photoionize atomic hydrogen. This creates a layer rich in atomic hydrogen as shown in Fig. \ref{fig:1star}C. Here, the north surface shows an increase in the abundance of atomic hydrogen. Immediately behind the ionization front (i.e. towards the interior of the cloud), the free hydrogen atoms transition to molecular hydrogen as shown in Fig. \ref{fig:1star}B. Figures \ref{fig:1star}D and E show the distribution of C~{\sc ii} and C~{\sc i} respectively. By comparing these two panels it can be seen how the C~{\sc ii} to C~{\sc i} transition occurs in the spheroid. Figure \ref{fig:1star}F plots the local emissivity of [C~{\sc i}] $609\,\mu{\rm m}$. This line is predominantly emitted from a thin region located at low optical depths i.e. close to the ionization front, and its peak is at $A_{\rm V}\sim1\,{\rm mag}$.

We emphasize that the H~{\sc i}-to-H$_2$ transition is known to occur in a very thin zone of the PDR and consequently in order to properly resolve it we need to increase the resolution of the grid. In the work presented here, none of the 3D simulations use high spatial resolution in order to study this transition in detail. As shown in \citet{Offn13}, such a low spatial resolution (i.e. not being able to resolve the H~{\sc i} to H$_2$ transition) does not change the overall chemistry of the PDR. However, by using the AMR techniques as described in Appendix \ref{app:tech}, {\sc torus-3dpdr} is able to perform such detailed studies at the cost of computational expense.

\subsection{Effect of the diffuse radiation (S2)}
\label{ssec:S2}

\begin{figure*}
  \includegraphics[width=0.95\textwidth]{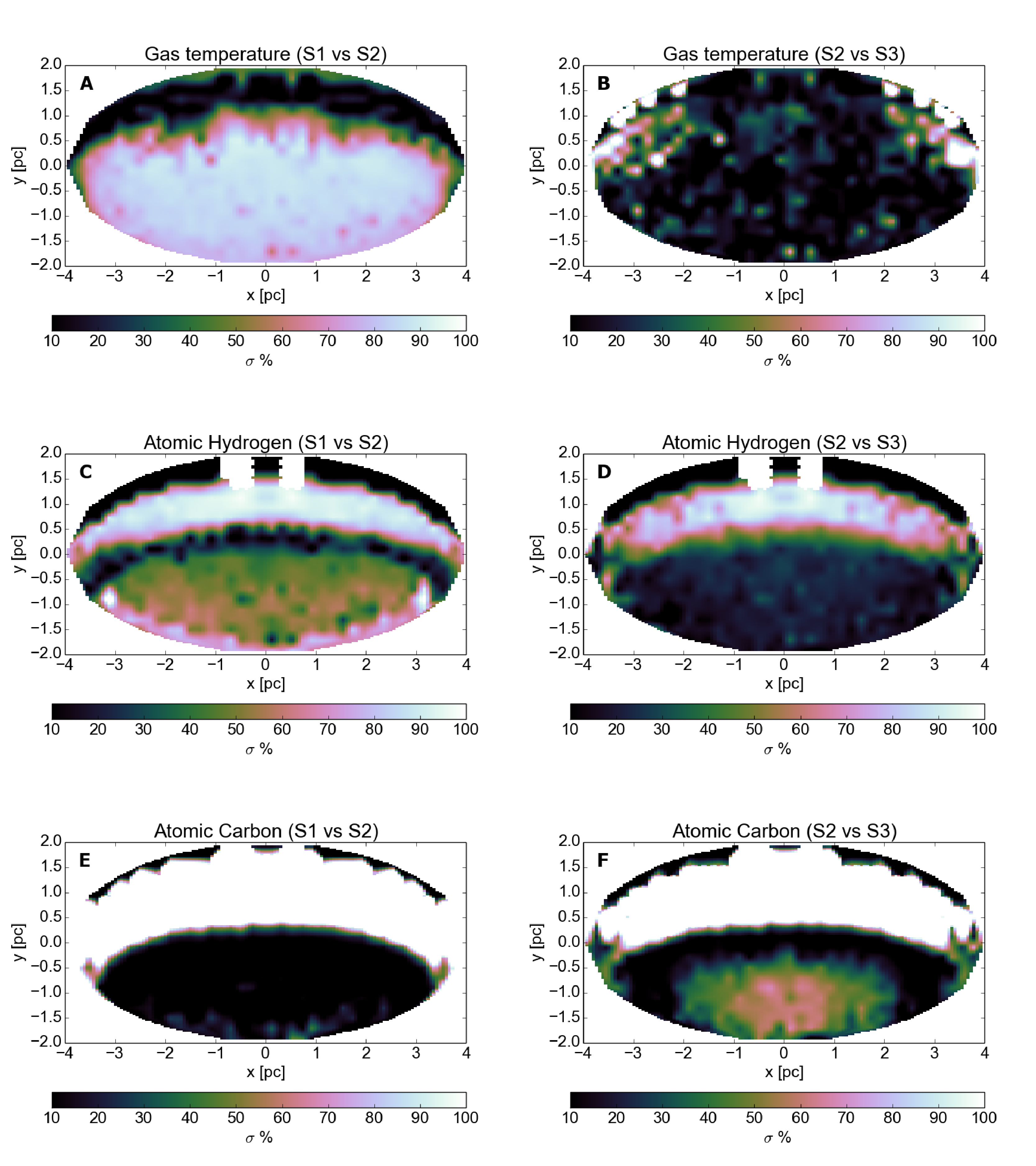}%S1S2S3ellipses.eps}
  \caption{ Cross section plots showing the relative change between S1 and S2 simulations (left column) and S2 and S3 simulations (right column). Thus the left column corresponds to the effects observed when including the diffuse component of the radiation field while the right column corresponds to when a multiple source treatment is taken into consideration. From top to bottom, we plot the relative change of the gas temperature, atomic hydrogen abundance, and atomic carbon abundance. The noise observed at the innermost part of the spheroid in Panel B (and which corresponds to a $\sim20\,{\rm K}$ temperature difference) is due to the thermal balance accuracy used. The $x-$ and $y-$ axes are in ${\rm pc}$. Relative changes $<5\%$ and $>100\%$ are not shown. As a result of the additional diffuse component of the radiation field in S2, the ellipsoid is warmer (by $\sim40\,{\rm K}$ on average) thus changing the abundances of species accordingly and particularly at low optical depths (left column). On the other hand, a multiple source treatment (right column) shows that although the gas temperature remains remarkably similar between the S2 and S3 simulations, H~{\sc i} and C~{\sc i} show a dependence on the UV field structure at low optical depths.}
  \label{fig:diff1star}
\end{figure*}

To explore the effect of diffuse radiation on the abundance distribution of different species, we repeat the S1 simulation in which we are additionally considering a full treatment of the UV field. The diffuse component of the radiation field penetrates further into the cloud, thus increasing its temperature and changing its chemistry and line emission. We focus here on the changes observed in the oblate spheroidal cloud itself between the S1 and S2 simulations, hence we will not include in our discussion the ambient medium. To show the effects of the interaction of the diffuse field we plot in the left column of Fig. \ref{fig:diff1star} cross sections of the spheroid, coloured according to the percentage of the relative change, $\sigma\%$, between the two simulations, given by the relation
\begin{eqnarray}
\sigma\%=\left|\frac{S_i-S_j}{S_j}\right|\cdot100\%
\end{eqnarray}
where $S_i$ and $S_j$ are the values of the $i$ and $j$ simulation that we compare.

The relative change in the gas temperature is shown in Fig. \ref{fig:diff1star}A. Here, although the temperature remains the same close to the ionization front (with a relative change of $<20\%$), it gets hotter in the inner part of the cloud (with a relative change of $\sim90\%$). In particular, at high optical depths we find that the spheroid in the S1 simulation has an average temperature of $T_{\rm gas}\sim10\,{\rm K}$, whereas in the S2 simulation it has $T_{\rm gas}\sim100-200\,{\rm K}$. As a consequence, the abundance of atomic hydrogen is higher in the S2 simulation reducing at the same time the abundance of H$_2$. The H~{\sc i}-to-H$_2$ transition has now been shifted further inside the cloud (but still at low optical depths; see Fig.~\ref{fig:diff1star}C). Similarly, C~{\sc i} is primarily emitted by a region closer to the centre of the ellipsoid. Since the peak of C~{\sc i} occurs in general in a quite thin zone of the PDR, such a shift may cause significant changes in its abundance distribution through the ellipsoid as observed in Fig.~\ref{fig:diff1star}E. However, in the innermost region, both S1 and S2 simulations are in excellent agreement. Hence inclusion of the diffuse component of the radiation field primarily affects the abundance distribution at low optical depths, while it is able to heat up the gas at intermediate optical depths.

\begin{figure*}
  \includegraphics[width=0.95\textwidth]{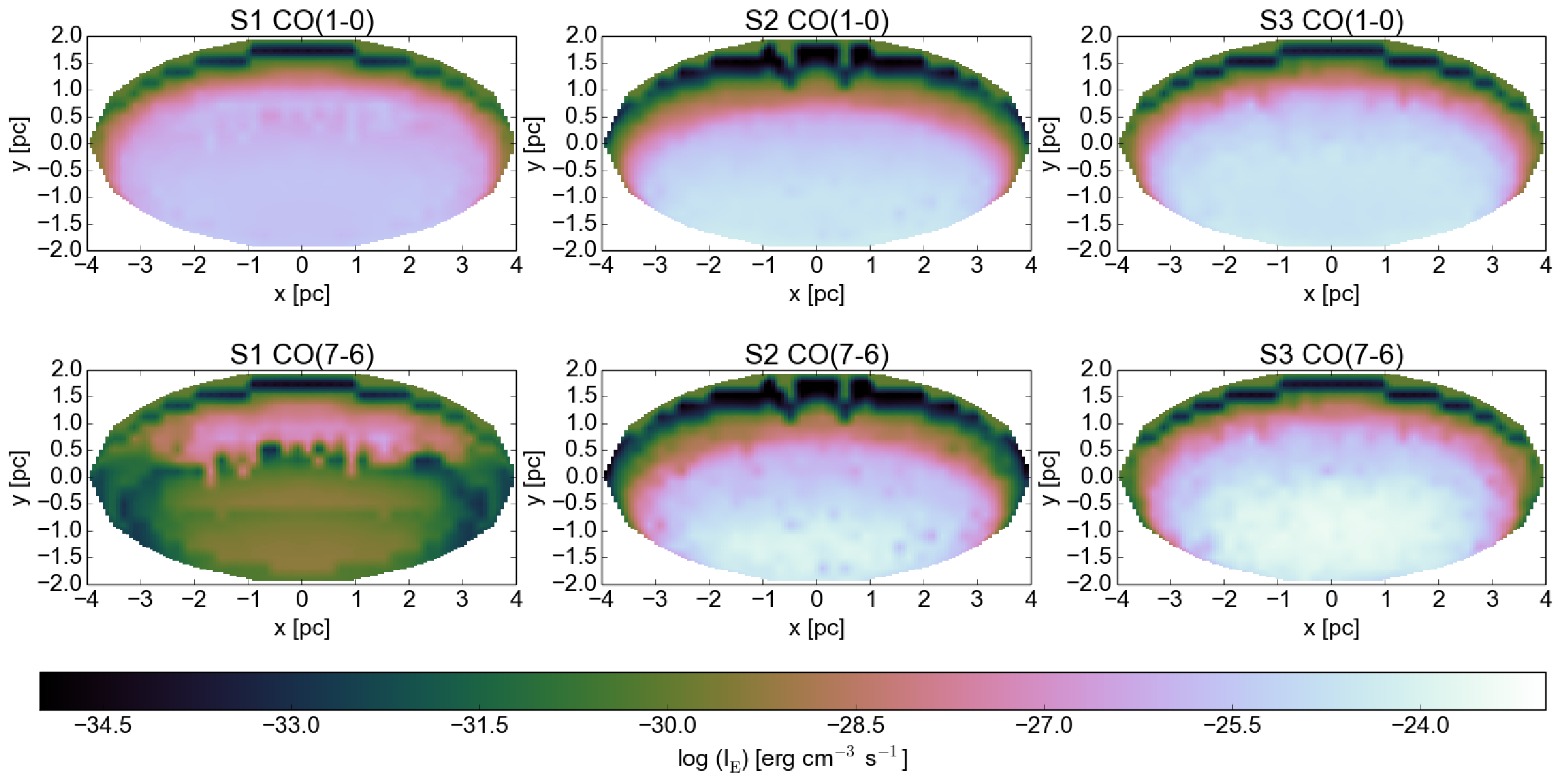}
  \caption{ Cross section plots showing local emissivities of CO for the transitions $J=1-0$ (upper row) and $J=7-6$ (lower row). The first column corresponds to the S1 simulation, the second column corresponds to the S2 simulation and the third one to the S3. The units of the logarithmic colour bar are in ${\rm erg}\,{\rm cm}^{-3}\,{\rm s}^{-1}$. The diffuse component of the radiation field heats up the inner part of the cloud by a few K which result in changing the emissivity of the different $J$ transition lines with the high ones to be more affected, as we observe by comparing the S1 and S2 simulations. However, a multiple source treatment (simulation S3) results in an emissivity map not significantly different from that of the S2 simulation, although it affects the emission at low optical depths (i.e. close to the ionization front).}
  \label{fig:highJ}
\end{figure*}

Figure \ref{fig:highJ} shows cross section plots of the ellipsoid where the local emissivity of CO $J=1-0$ and CO $J=7-6$ transition lines are mapped (left and middle panels for the S1 and S2 simulations respectively). The difference in gas temperature at the inner part of the cloud due to the diffuse component of radiation leads to discrepancies in the emissivities of both $J$ transition lines with the high-$J$ to be up to $\sim5$ orders of magnitude brighter in the S2 simulation. This shows that high-$J$ CO lines may be used as gas temperature diagnostics in regions of intermediate optical depths where the diffuse radiation can penetrate the cloud deeper. It also shows that a realistic treatment of the UV field is necessary when comparing observational with numerical data such as i.e. of the Orion Bar \citep[i.e.][]{Tiel93,vdWe96,Andr14}.
%Table \ref{tab:average} shows the average values of the gas temperature and the abundances of H$_2$, H~{\sc i}, C~{\sc ii}, C~{\sc i} and CO. These have been obtained by summing up the values of all individual PDR cells of the ellipsoid and then by dividing with the total number of those cells. By comparing the S1 and S2 columns we see that when including the diffuse field, $<T_{\rm gas}>$ is $\sim2$ times higher than when by invoking the OTS approximation. By comparing the average values of the abundances however, we see that only the CO abundance is quite similar in both S1 and S2 since the diffuse component has affected H$_2$, H~{\sc i}, C~{\sc ii} and C~{\sc i} at lower optical depths.
%
%
%\begin{table}
% \centering
%  \caption{Average values for gas temperature and abundances of H$_2$, H~{\sc i}, C~{\sc ii}, C~{\sc i} and CO for the S1 (first column), S2 (middle column) and S3 (right column) simulations. These correspond only to the ellipsoid thus neglecting the contribution of the ambient medium.}
%  \label{tab:average}
%
%  \begin{tabular}{l c c c}
%  \hline
%   Quantity & S1 (OTS approximation) & S2 (Diffuse field) & S3 (2 stars) \\
%     \hline
%   T$_{\rm gas}$ [K] & 43 & 83 & 89 \\
%   H$_2$ & 0.44 & 0.39 & 0.44 \\
%   H~{\sc i} & 0.12 & 0.21 & 0.12 \\
%   C~{\sc ii} & $1.16\times10^{-4}$ & $1.4\times10^{-4}$ & $1.2\times10^{-4}$ \\
%   C~{\sc i} & $9.7\times10^{-5}$ & $6.9\times10^{-5}$ & $7.7\times10^{-5}$ \\
%   CO & $7.3\times10^{-6}$ & $7\times10^{-6}$ & $1.9\times10^{-5}$ \\
% \hline
%\hline
%\end{tabular}
%\end{table}

\subsection{Effect of multiple sources (S3)}
\label{ssec:S3}

The effect of multiple sources interacting with PDRs is particularly interesting and very poorly studied due to the lack of numerical codes that are able to treat such systems. Here, we discuss the capabilities of {\sc torus-3dpdr} in modelling such PDRs by revisiting the S2 simulation described above, in which we replace the ionizing source of $\dot{\cal N}_{\rm LyC}=1.2\times10^{50}\,{\rm s}^{-1}$ with two identical sources of $\dot{\cal N}'_{\rm LyC}=6\times10^{49}\,{\rm s}^{-1}$. Both sources in this simulation have the same distance $D$ from the center of the oblate spheroidal cloud as in the S1 and S2 simulations. The diffuse component of the radiation field is taken into consideration, hence this setup ensures that any changes observed between S2 and S3 will be a result of the structure of the radiation field emitted by two sources instead of a single source. 

The right panel of Fig. \ref{fig:diff1star} shows the relative change between S2 and S3 simulations for the gas temperature, the atomic hydrogen abundance, and the atomic carbon abundance. We find that the gas temperature shows fluctuations of the order of $<30\%$ for the entire cloud, corresponding to $\sim20\,{\rm K}$ (see Fig. \ref{fig:diff1star}B).
%We find that the gas temperature is very little affected by the multiple source treatment (with average relative change $<30\%$ for the entire cloud). The noise observed in Fig. \ref{fig:diff1star}B corresponds to a difference of $\lesssim20\,{\rm K}$ in temperature. Although the gas temperature remains unaffected, 
The spatial distribution of the UV field has an impact in the abundances distribution between S2 and S3 particularly at low optical depths. This can be seen in Fig.~\ref{fig:diff1star}D and \ref{fig:diff1star}F, where we compare H~{\sc i} and C~{\sc i}. Similarly to what has been discussed in \S\ref{ssec:S2}, the UV distribution is now less strong as in S2 resulting to a shift of the H~{\sc i}-to-H$_2$ transition to lower optical depths, thus also increasing H$_2$. Such a behaviour is also observed in C~{\sc i}. 

\begin{figure*}
  \includegraphics[width=0.95\textwidth]{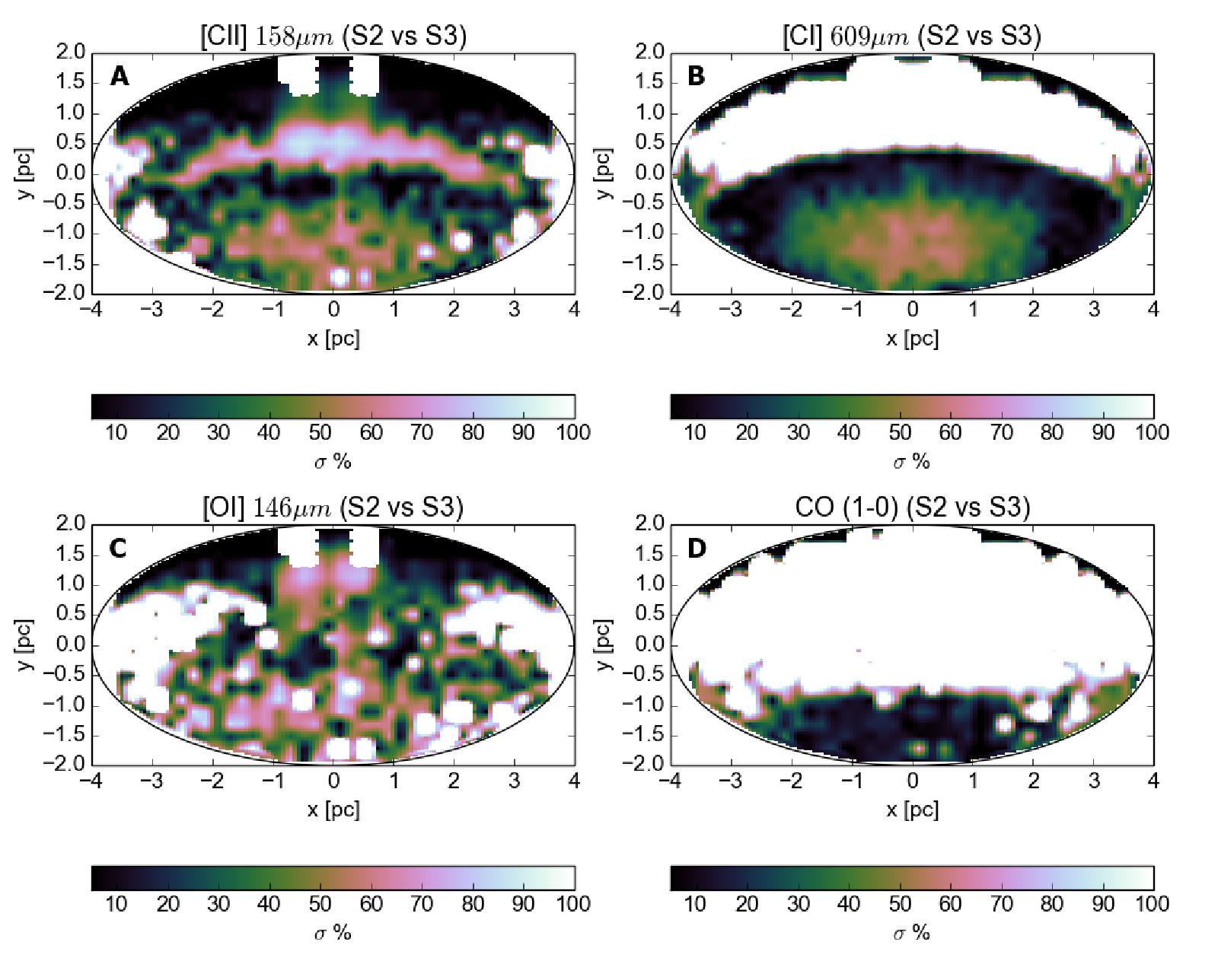}%S2S3ellipsediff.eps}
  \caption{ Cross section plots showing the relative change of the local emissivities for various cooling lines for the S2 and S3 simulations and where in both cases diffuse radiation has been taken into consideration. Both axes are in ${\rm pc}$. Panel A shows the relative change for the [C~{\sc ii}] $158\mu m$ line. Panel B for the [C~{\sc i}] $609\mu m$, panel C for [O~{\sc i}] $146\mu m$ and panel D for CO (1-0). From these plots we see that overall multiple sources have an impact on the emissivities of the cloud, although in both S2 and S3 simulations the total intensity of UV radiation used was the same. As explained also in Fig.\ref{fig:diff1star}, the areas close to ionization front and to the $x=0\,{\rm pc}$ axis of symmetry are subject to spurious heating due to a small number of high energy photons.}
  \label{fig:difflines}
\end{figure*}

%In Table~\ref{tab:average} we compare the average values of S2 and S3. We observe here that the gas temperatures remain remarkably similar, but the abundances of H$_2$ and H~{\sc i} are more similar to S1. It is interesting to note that in the S3 simulation, the abundance of CO is the highest when compared to both S1 and S2. 
The middle and right columns of Fig. \ref{fig:highJ} compares CO $J=1-0$ and CO $J=7-6$ for the S2 and S3 simulations. In both cases the differences observed are not significant in comparison with S1 and S2, however at the inner part of the cloud the $J=7-6$ line emissivity is higher by a factor of $\sim2$ reflecting the gas temperature difference of $\sim20\,{\rm K}$ between S2 and S3. We argue that in more complicated structures involving multiple sources the effects introduced by the realistic UV treatment can be of significant importance. Furthermore, we plot in Fig. \ref{fig:difflines} the relative changes in the line emissivities for the S2 and S3 simulations of C~{\sc ii} $158\,\mu{\rm m}$, C~{\sc i} $609\,\mu{\rm m}$, O~{\sc i} $146\,\mu{\rm m}$ and CO $J=1-0$. In general, C~{\sc ii} $158\,\mu{\rm m}$ remains in principle unaffected when the UV radiation field structure is taken into account, however C~{\sc i} $609\,\mu{\rm m}$ and in particular CO $J=1-0$ show strong dependency under a multiple source treatment. These point to the conclusion that emission lines do become dependent on the spatial structure of the radiation field even when the total intensity received is the same as the intensity emitted by a single source at the same distance. This is in agreement with the results presented in B12 as well as with the findings of \citet{Offn13}. Thus three-dimensional codes such as {\sc torus-3dpdr} are often needed in order to compare as closely as possible numerical simulations with observations.

%%%%%%%%%%%%%%%%%%%%%%
\section{Conclusions}%
%%%%%%%%%%%%%%%%%%%%%%
\label{sec:conclusions}

In this paper we present {\sc torus-3dpdr}, a three-dimensional astrochemistry code which is able to treat simultaneously photoionization and photodissociation regions with arbitrary geometrical and density distributions interacting with single or multiple sources. The code uses a Monte Carlo-based scheme to calculate the propagation of the ionizing radiation including its diffusive component and a {\sc healpix}-based ray-tracing scheme to calculate the column densities in each cell and along any direction on the celestial sphere. We have performed one-dimensional and three-dimensional simulations demonstrating the capabilities of the code. 

We performed a one-dimensional PDR calculation to explore the consistency of {\sc torus-3dpdr} against {\sc 3d-pdr} \citep{Bisb12} and we found excellent agreement between the two codes in reproducing the temperature profile as well as the abundances of species and the local emissivities of the major cooling lines. The simulation setup was identical to a standard PDR benchmarking test discussed in \citet{Roll07}, although our codes include the updated UMIST network as well as an updated H$_2$ formation heating mechanism and thus we can not compare {\sc torus-3dpdr} directly against the numerous codes presented in \citet{Roll07}. 

We then advanced our benchmarking calculation by performing a one-dimensional calculation including both the MC photoionization treatment as well as the PDR treatment in a uniform density cloud. The photoionization calculations reproducing the H~{\sc ii} region followed the Lexington HII40 benchmarking case and showed excellent agreement with the {\sc mocassin} code \citep{Erco03, Wood04, Erco05}. Once the value of the ionization fraction of hydrogen fell below $\chi_{\rm i}<0.3$, corresponding to a gas temperature of $T\sim3000\,{\rm K}$, the PDR calculations were switched on. The thermal balance obtained in the PDR using {\sc torus-3dpdr} was in agreement with {\sc 3d-pdr} only when we included the extinction due to the distance of each PDR cell for the ionizing radiation emitted by the ionizing point source. We thus argue that this factor is important in order to obtain more accurate abundance distributions in the PDR and particularly a more accurate gas temperature profile. 

We have demonstrated the three-dimensional capabilities of {\sc torus-3dpdr} by treating an application in which we considered an oblate spheroidal cloud irradiated externally by either a single source or two sources. Although this simulation setup is simplified in three-dimensional space, it is not reproducible by one-dimensional codes. We have performed control runs in which we have invoked the on-the-spot approximation and additionally the effects introduced by the diffuse component of radiation field. Our results for this application can be summarized as follows:

\begin{enumerate}
\item The effects introduced by the diffuse component of the radiation field in the single ionizing source case (simulations S1 and S2), have been found to affect both the outer parts (low optical depths) and the inner parts (high optical depth) of the cloud. We found that the gas temperature has been increased as a result of the diffuse radiation by an average of $\sim40\,{\rm K}$ and by $\sim100-200\,{\rm K}$ individually at high optical depths. This in turn resulted in a higher abundance of H~{\sc i} while also increasing locally the abundance of C~{\sc ii} compared to what we found when invoking the on-the-spot approximation (S1). 
\item The effects introduced by the interaction of two sources instead of a single source (simulations S2 and S3) can heat up the gas at the inner part of the cloud by $\sim20\,{\rm K}$ which results to a $\sim2$ times higher line emissivity of the CO $J=7-6$ transition. We argue that the impact of multiple source treatment is potentially higher at even higher CO $J$ transitions. In addition, the abundance distributions of the species examined are strongly correlated with the geometric distribution of the radiation field.
\item The local emissivities of the fine-structure lines which are predominantly emitted at low optical depths, depend on the spatial shape of the radiation field (simulations S2 and S3). The local emissivities of CO transitions are strongly affected, showing that we require realistic and detailed UV structures when comparing models with observational data.
\end{enumerate}

In addition, we have performed a test in which we increased the angular resolution ({\sc healpix} level of refinement) in the S1 simulation and examined the gas temperature structure and H$_2$ abundance distribution. We found that even at the lowest possible resolution ($\ell=0$ which also compromises a low computational cost) {\sc torus-3dpdr} provides acceptable results for the simulations presented in this paper. For more complicated density distributions however, a higher level of {\sc healpix} refinement is needed though at the cost of computational expense.

While most existing PDR codes consider one-dimensional density profiles and very simplified radiation fields, {\sc torus-3dpdr} brings us a step closer to simulating realistic three-dimensional H~{\sc ii}/PDR complexes interacting with multiple sources. Future versions of {\sc torus-3dpdr} will include treatments of dust radiation transfer and X-ray chemistry in the photoionization and photodissociation regions.

%%%%%%%%%%%%%%%%%%%%%%%%%%%%%%%%%%%%%%%%%%%%%%%%%%%%%%%%%%%%%5
%%%%%%%%%%%%%%%%%%%%%%%%%%%%%%%%%%%%%%%%%%%%%%%%%%%%%%%%%%%%%5
\section*{Acknowledgments}
The authors thank the anonymous referee for the useful comments which have significantly improved the validity of the {\sc torus-3dpdr} code. The work of TGB was funded by STFC grant ST/J001511/1. The work of TJHaw was funded by STFC grant ST/K000985/1. TGB and TJHaw acknowledge the NORDITA programme on Photo-Evaporation in Astrophysical Systems (2013 June) where part of the work for this paper was carried out. This research has made use of NASA's Astrophysics Data System.

Part of this work used the DiRAC Data Analytics system at the University of Cambridge, operated by the University of Cambridge High Performance Computing Serve on behalf of the STFC DiRAC HPC Facility (www.dirac.ac.uk). This equipment was funded by BIS National E-infrastructure capital grant (ST/K001590/1), STFC capital grants ST/H008861/1 and ST/H00887X/1, and STFC DiRAC Operations grant ST/K00333X/1. DiRAC is part of the National E-Infrastructure.

%%%%%%%%%%%%%%%%%%%%%%%%%%%%%%%%%%%%%%%%%%%%%%%%%%%%%%%%%%%%%5
%%%%%%%%%%%%%%%%%%%%%%%%%%%%%%%%%%%%%%%%%%%%%%%%%%%%%%%%%%%%%5

\appendix

\section{Technical characteristics}
\label{app:tech}

\subsection{Ray tracing on a domain decomposed grid}
{\sc torus-3dpdr} uses the octree Adaptive Mesh Refinement (AMR) grid native to {\sc torus}, including domain decomposition. Domain decomposition splits the grid into subsets, each of which are handled by processors with independent private memory. This provides the advantage of alleviating intensive memory requirement of of {\sc 3d-pdr}. However, during ray tracing, rays must traverse cells on the grid that belong to different domains to that from which the ray originate, in which the thread doing the ray tracing has no knowledge of the conditions. To overcome this issue, Message Passing Interface (MPI) communication is required between the  thread doing the ray tracing and the thread through which the ray is passing. To facilitate this, we iterate over domains, with one performing ray tracing operations and the others acting as servers. MPI does not therefore offer significant speedup in the ray tracing component of the calculations. It does however offer speed up in other components of the PDR calculation and offers excellent speedup for the Monte Carlo radiative transfer part of the calculation \citep{Harr15}.

\subsection{On-the-fly calculation}
As explained in \citet{Bisb12}, the {\sc 3d-pdr} code stores information on every intersection point for every ray. Although this offers the ability to perform PDR calculations directly on any SPH or grid-based simulation without further modifications, it has disadvantages with regards to the large memory requirements. In the new coupled code, we perform calculations on a cell-by-cell basis. We take advantage of the fact that {\sc torus} moves recursively through the AMR grid throughout the calculations.

\subsection{Hybrid parallelization}
\label{app:cputime}
In addition to the domain decomposition we simultaneously parallelize the code using Open Multi-Processing (OpenMP) over intensive loops in the calculation. On each computational node, one (or more) cores are assigned the role of MPI threads and the others are used for OpenMP parallelization. Use of MPI and OpenMP in this hybrid manner provides us with a flexible parallelization scheme that both reduces calculation wall time and memory demands. For the tests performed here, we find that the computational expense depends strongly on each different application. The one-dimensional HII/PDR calculation presented in \S\ref{ssec:photopdr}, required approximately 22 CPU minutes. The PDR calculations in the three-dimensional (excluding the Monte Carlo photoionization part) required $\sim148$ CPU hours for $\ell=0$ {\sc healpix} refinement, $\sim522$ CPU hours for $\ell=1$ and $\sim1251$ CPU hours for $\ell=2$. These show that an $\ell=2$ simulation is about an order of magnitude more expensive than one at $\ell=0$. We are currently upgrading the OpenMP parallelization and with recent developments by \citet{Harr15}, the computational cost will be reduced in future versions of {\sc torus-3dpdr}. These further optimizations will make computationally more tractable complicated post processing of 3D simulations.

\section{Flowchart of {\sc torus-3dpdr}}
\label{app:flow}

\begin{figure}
\psfrag{Start}{Start {\sc torus-3dpdr}}
\psfrag{Propagate}{Propagate photons}
\psfrag{Ionization}{Ionization balance}
\psfrag{Thermal}{Thermal balance}
\psfrag{xCHEMITER}{ITERATE}
\psfrag{Converge}{Is it converged?}
\psfrag{RayTracing}{HEALPix ray casting}
\psfrag{Do3DPDR}{{\sc 3d-pdr} as in \citet{Bisb12}}
\psfrag{End}{End {\sc torus-3dpdr}}
\psfrag{True}{\texttt{TRUE}}
\psfrag{False}{\texttt{FALSE}}
\psfrag{PDR}{PDR}
\psfrag{PhotoI}{Photoionization}
\includegraphics[width=0.5\textwidth]{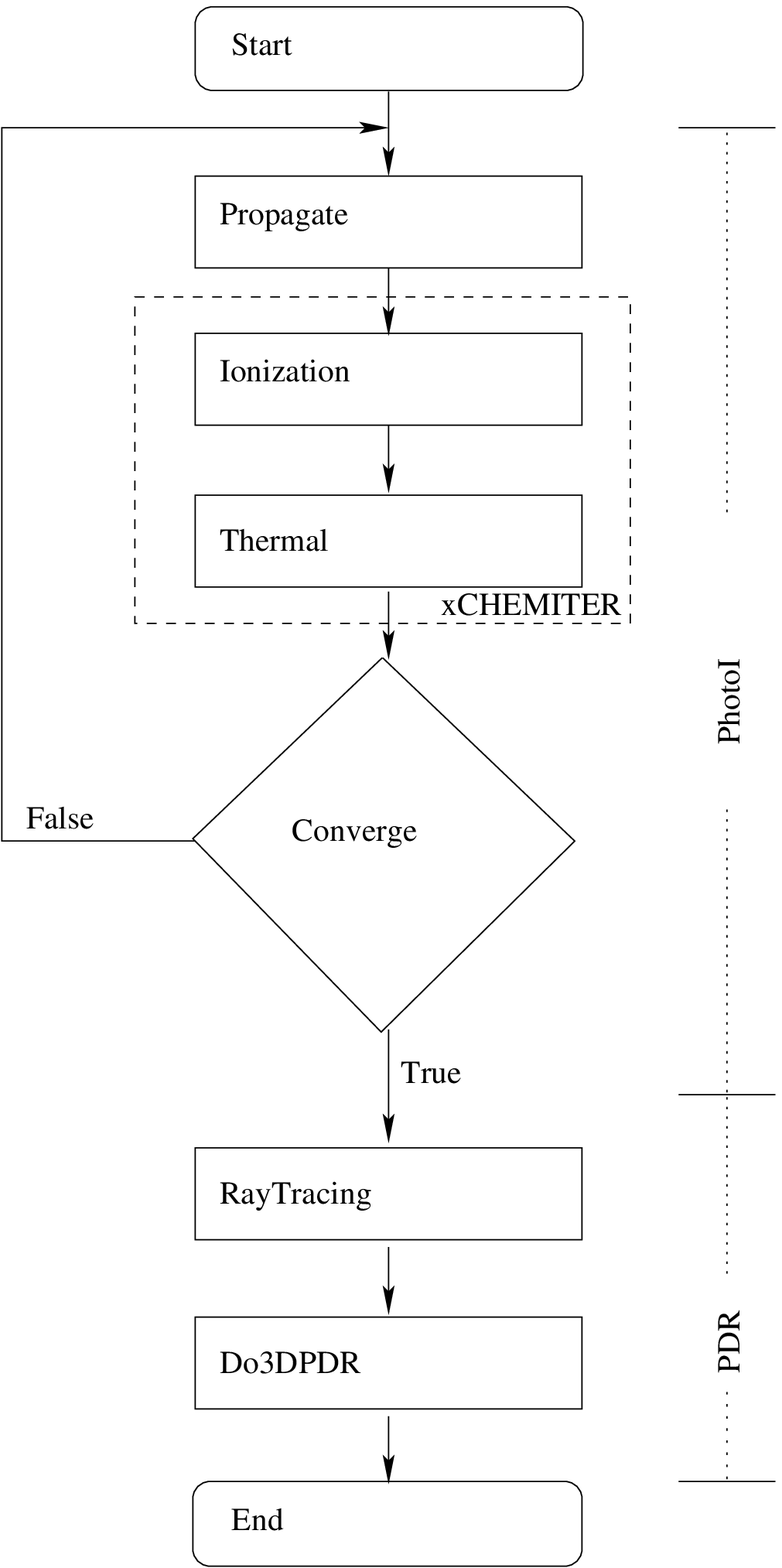}
\caption{ Flowchart of {\sc torus-3dpdr}}
\label{fig:flowchart}
\end{figure}

Figure \ref{fig:flowchart} shows the basic flowchart describing how {\sc torus-3dpdr} converges in evaluating the photoionization (upper half of the Figure) and PDR (lower half of the Figure) calculations. Each solid box represents a DO-loop over grid cells. The dashed box represents iterations over the ionization and the thermal balance in the photoionization calculations.

The code starts by reading all input model parameters, such as the density distribution, the abundances of species, and the location and properties of each ionizing source. It then starts the Monte Carlo based propagation of ionizing photons and it iterates in order to obtain equilibrium in the H~{\sc ii} region. Once {\sc torus-3dpdr} reaches photoionization convergence, it stores the direction of the UV radiation field in each cell. The code then proceeds further in performing the PDR calculations. In particular {\sc healpix}-based rays emanate from each cell flagged as PDR, and it moves along them to calculate the column densities for every cell in every direction. A cell is flagged as PDR if the value of the hydrogen ionization fraction is below $\chi_{\rm i}<0.3$, corresponding to a gas temperature of $T\sim3000\,{\rm K}$. 

Any further PDR calculation follows the flowchart presented in Appendix A of \citet{Bisb12}. Once {\sc torus-3dpdr} is fully converged it writes the outputs and terminates.

\section{Geometrical dilution factor}
\label{app:dilution}

The flux, $F$, at a given point point $R$ is given by the equation
\begin{eqnarray}
\label{eqn:app.1}
F=\frac{L_{0}}{4\pi R^2},
\end{eqnarray}
where $L_{0}$ is the luminosity of the source (in this case a single star). At the position of the ionization front (IF), the above equation leads to
\begin{eqnarray}
\label{eqn:app.2}
L_{0}=4\pi R_{\rm IF}^2 F_{\rm IF}.
\end{eqnarray}
At a given position, $r$, inside the PDR (hence at a distance $R=R_{\rm IF}+r$ from the star), Eqn.(\ref{eqn:app.1}) takes the form
\begin{eqnarray}
\label{eqn:app.3}
F_{\rm r}=\frac{L_{0}}{4\pi (R_{\rm IF}+r)^2},
\end{eqnarray}
which when combined with Eqn.(\ref{eqn:app.2}) leads to
\begin{eqnarray}
\label{eqn:app.4}
F_{\rm r}=F_{\rm IF}\frac{R_{\rm IF}^2}{(R_{\rm IF}+r)^2}.
\end{eqnarray}
From the above it turns that the factor ${R_{\rm IF}^2}/{(R_{\rm IF}+r)^2}$ corresponds to the geometrical dilution of the UV radiation when account is taken of the dilution inside the H~{\sc ii} region. Hence, if we additionally take into account the exctinction of the radiation along distance $r$, it leads to Eqn.(\ref{eqn:geomdil}). Neglecting the dilution of the radiation inside the H~{\sc ii} region, the corresponding factor is simply $(4\pi r^2)^{-1}$, leading to Eqn.(\ref{eqn:uvtest}). We note that a full treatment of the H~{\sc ii}/PDR complex using the MC method takes into account the dilution inside the ionized region, hence it is in agreement with Eqn.(\ref{eqn:geomdil}). In addition, for simplified cases such as in the plane-parallel approximation, {\sc torus-3dpdr} is able to treat the UV field as discussed in \citet{Roll07}. 

%%%%%%%%%%%%%%%%%%%%%%%%%%%%%%%%%%%%%%%%%%%%%%%%%%%

\newcommand{\apj}[1]{ApJ, }
\newcommand{\apss}[1]{Ap\&SS, }
\newcommand{\aj}[1]{Aj, }
\newcommand{\apjs}[1]{ApJS, }
\newcommand{\apjl}[1]{ApJ Letter, }
\newcommand{\aap}[1]{A\&A, }
\newcommand{\aaps}[1]{A\&A Suppl. Series, }
\newcommand{\araa}[1]{Annu. Rev. A\&A, }
\newcommand{\aaas}[1]{A\&AS, }
\newcommand{\bain}[1]{Bul. of the Astron. Inst. of the Netherland,}
\newcommand{\mnras}[1]{MNRAS, }
\newcommand{\nat}[1]{Nature, }
\newcommand{\araaa}[1]{ARA\&A, }
\newcommand{\planss}[1]{Planet Space Sci., }
\newcommand{\jrasc}[1]{Jr\&sci, }
\newcommand{\pasj}[1]{PASJ, }
\newcommand{\pasp}[1]{PASP, }

\end{document}